\documentclass[journal]{IEEEtran}
\ifCLASSINFOpdf

\else

\fi

\usepackage{graphicx,amsmath,amssymb,cite}
\usepackage{color}
\usepackage{amsfonts}
\usepackage{amsmath}
\usepackage{bm}
\usepackage{graphicx}
\usepackage{amssymb}
\usepackage{stfloats}
\usepackage{mathrsfs}
\usepackage{mathtools} 
\usepackage{subfloat}  
\usepackage{subcaption} 
\usepackage{tabularx}
\usepackage{booktabs}

\begin{document}
\title{DNN-based Methods of Jointly Sensing Number and Directions of Targets via a Green Massive $\text{H$\mathrm{^2}$AD}$ MIMO Receiver}
	\author{Bin Deng, Jiatong Bai, Feilong Zhao, Zuming Xie, Maolin Li, Yan Wang, Feng Shu
    \thanks{(Corresponding author: Feng Shu.)}
	\thanks{Bin Deng, Jiatong Bai, Yan Wang, and Maolin Li are with the School
		of Information and Communication Engineering, Hainan University, Haikou,
		570228, China (E-mail: d2696638525@126.com, 18419229733@163.com, yanwang@hainanu.edu.cn, limaolin0302@163.com).}
	\thanks{Feng Shu is with the School of Information and Communication Engineering and Collaborative Innovation Center of Information Technology, Hainan University, Haikou 570228, China, and also with the School of Electronic and Optical Engineering, Nanjing University of Science and Technology, Nanjing 210094, China (E-mail: shufeng0101@163.com).}
	\thanks{Feilong Zhao and Zuming Xie are with the China Mobile Group Hainan Co., Ltd., Haikou 571250, China (E-mail: somethingnothing@163.com, xiezuming@hi.chinamobile.com).}}

\maketitle

\begin{abstract}
As a green MIMO structure, the heterogeneous hybrid analog-digital ($\text{H$\mathrm{^2}$AD}$) MIMO architecture has been shown to own a great potential to replace the massive or extremely large-scale fully-digital MIMO in the future wireless networks to address the three challenging problems faced by the latter: high energy consumption, high circuit cost, and high complexity. However, how to intelligently sense the number and direction of multi-emitters via such a structure is still an open hard problem. To address this, we propose a two-stage sensing framework that jointly estimates the number and direction values of multiple targets. Specifically, three target number sensing methods are designed: an improved eigen-domain clustering (EDC) framework, an enhanced deep neural network (DNN) based on five key statistical features, and an improved one-dimensional convolutional neural network (1D-CNN) utilizing full eigenvalues. Subsequently, a low-complexity and high-accuracy DOA estimation is achieved via the introduced online micro-clustering (OMC-DOA) method. Furthermore, we derive the Cramér–Rao lower bound (CRLB) for the H²AD under multiple-source conditions as a theoretical performance benchmark. Simulation results show that the developed three methods achieve 100\% number of targets sensing at moderate-to-high SNRs, while the improved 1D-CNN exhibits superior under extremely-low SNR conditions. The introduced OMC-DOA outperforms existing clustering and fusion-based DOA methods in multi-source environments.

\end{abstract}

\section{INTRODUCTION}



In wireless communications, radar systems, and autonomous unmanned systems, target localization and detection are fundamental to situational awareness and mission execution. To achieve accurate spatial localization of targets, several techniques have been widely adopted, including time of arrival (TOA), received signal strength (RSS), and direction of arrival (DOA)~\cite{WOS:000366848200002, 8290952, WOS:000424945600039}. In this context, UAV-based localization methods have been explored, but challenges like signal strength fluctuation remain~\cite{10644103}. Among these, DOA estimation stands out by leveraging the phase differences or power distributions of signals received at an antenna array to achieve high-precision angle localization with relatively limited resources. This makes DOA particularly suitable for modern communication scenarios such as large-scale MIMO~\cite{WOS:000777133000002}, cooperative sensing, and integrated sensing and communication (ISAC)~\cite{WOS:001442866900023, WOS:001284774200001}, as well as IRS and HIRS enhanced localization in complex propagation environments~\cite{10654551,10852155,LI2025110911}.

To  accommodate complex and dynamic application environments, existing DOA estimation methods can be broadly categorized into four types. 
The first category comprises conventional approaches such as beamforming~\cite{WOS:000811551300001}, maximum likelihood estimation (MLE)~\cite{WOS:001098745400007}, and fast fourier transform (FFT)~\cite{WOS:000399952000022}, which are easy to implement but perform poorly when resolving closely spaced sources.
The second category comprises subspace-based methods such as MUSIC~\cite{WOS:000364855000018} and ESPRIT~\cite{WOS:000438743000005}, which rely on eigen-decomposition of the covariance matrix. These methods offer excellent resolution performance. However, it still faces challenges, such as difficulties in estimating the number of signal sources and sensitivity to noise. To address these issues, several improved methods have been proposed~\cite{WOS:000947280000012, WOS:000329056200001, WOS:000497864700075, WOS:000856989500001}. The third category includes compressive sensing (CS) based approaches, which enhance resolution under sparse modeling assumptions and are well-suited for under-sampled conditions~\cite{1982A}. The final category comprises machine learning–based methods, which offer notable advantages such as data-driven modeling and strong adaptability, making them well-suited for complex scenarios involving nonlinearities and unknown system models. They have been successfully applied to improve the accuracy and robustness of DOA estimation, as demonstrated in~\cite{WOS:001203463300014, 10892212,WOS:000553167900002}.

Existing DOA estimation algorithms demonstrate strengths across various design paradigms. However, their performance is limited by specific array configurations and underlying system architectures. In large-scale array deployments and resource-constrained joint communication and sensing (JCAS) systems, achieving high-resolution multi-source localization under limited hardware conditions remains a critical challenge for the practical implementation of DOA techniques. To alleviate the hardware complexity induced by costly fully digital arrays, the hybrid analog-digital (HAD) architecture has been proposed as a low-complexity alternative solution~\cite{WOS:000372617600016, WOS:000376097300001}. This architecture pre-combines subarray beams in the analog domain by connecting multiple antennas to a single RF chain, thereby effectively reducing the number of RF links and ADC modules, and significantly lowering system power consumption and implementation cost~\cite{WOS:000402731600011, WOS:000895081000068, WOS:000838680900015, WOS:000452623700047}. In the typical sub-connected HAD architecture, each subarray samples signals from a fixed direction, and the final estimation is performed centrally in the digital domain. Several studies have proposed various DOA estimation methods tailored to this architecture. For instance,~\cite{WOS:000876768200084} proposed a low-complexity beam scanning algorithm based on submatrix multiplication (BSASM), which reconstructs the spatial covariance matrix with reduced computational cost to achieve high-accuracy DOA estimation. In~\cite{WOS:000870308700022}, a partially connected hybrid beamforming method was proposed to minimize the Cramér–Rao bound (CRB), aiming to enhance DOA estimation performance in multi-user ISAC systems. 
Nevertheless, despite its hardware efficiency, the HAD architecture also faces challenges. 
On one hand, the large inter-subarray spacing may lead to phase ambiguity. On the other hand, it requires at least multiple sampling time slots to complete a single DOA measurement, making it difficult to meet the single-slot and low-latency requirements of 6G~\cite{WOS:000833178300001, WOS:001157855100008}, where efficient, secure and intelligent surface-aided communication paradigms are becoming increasingly important~\cite{10666006,11028902}.

To address these challenges, the $\text{H$\mathrm{^2}$AD}$ architecture has attracted growing attention~\cite{10767772}. This architecture integrates multiple conventional homogeneous HAD modules and inherently eliminates phase ambiguity in DOA estimation. It offers signal processing latency comparable to that of fully digital architectures, while maintaining hardware cost and power consumption similar to traditional HAD systems, thereby achieving a better trade-off between performance and complexity.
Recent work integrates fully‑digital modules with the $ \text{H$\mathrm{^2}$AD} $ architecture to boost DOA accuracy via co‑learning and derives its CRLB for performance benchmarking~\cite{2024arXiv240509556B}.
However, a critical yet often overlooked issue in DOA estimation is the lack of prior knowledge of the number of sources. Most classical DOA algorithms, such as MUSIC and root-MUSIC, either assume a known number of sources or are limited to single-source scenarios. This assumption is impractical for applications such as speech source localization~\cite{WOS:001089305500015}, cooperative UAV sensing~\cite{WOS:001252619600143}, and multi-target dynamic tracking~\cite{WOS:001128031700048}. Directly applying DOA estimation without distinguishing the number of sources not only leads to inefficient use of computational resources but also increases the risk of sensing failure. This issue becomes particularly critical in $\text{H$\mathrm{^2}$AD}$ systems, where the number of RF chains is limited, array structures are complex, and sensing resources are highly constrained. Blind multi-source searching over the full angular space can substantially increase system latency and energy consumption. Moreover, the superposition of multiple source signals disrupts phase consistency among subarrays, causing the noise subspace to deviate from ideal orthogonality and leading to complete failure of traditional methods such as root-MUSIC under multi-source conditions.

Motivated by these observations, this paper investigates a two-stage multi-source signal sensing framework tailored for large-scale $\text{H$\mathrm{^2}$AD}$ MIMO receivers. The proposed framework jointly addresses the problems of source number estimation and DOA estimation. The main contributions of this work are summarized as follows:
\begin{enumerate}
  \item To accurately estimate the number of targets, we propose an improved eigen-domain clustering (EDC) framework that exploits the separability between signal and noise eigenvalues obtained from the eigendecomposition of the covariance matrix. Since the steering vectors of the signals form the eigenvectors of the signal subspace, there exists an inherent correspondence between the signal subspace and the directions. The signal eigenvalues carry energy-related information, and their count is closely related to the number of sources. By setting appropriate thresholds or applying clustering techniques, the signal and noise eigenvalues can be effectively separated. Based on this, a density-based clustering (DBSCAN) algorithm is applied in the transformed eigenvalue space to estimate the number of targets in an unsupervised manner. The proposed method demonstrates accurate performance even under low-SNR and heterogeneous array conditions.

  \item To overcome the sensitivity of the improved EDC to cluster boundary ambiguity under extremely-low SNR regions, an enhanced deep neural network (DNN) is designed. By incorporating spectral entropy, which quantifies the uniformity of energy distribution among the eigenvalues of the covariance matrix, into conventional statistical features, the feature discriminability across different source counts is significantly enhanced. Moreover, a improved one-dimensional convolutional neural network (1D-CNN) is further proposed,which directly takes the normalized full eigenvalue sequence as input to perform end-to-end targets number estimation. Simulation results show that all three methods achieve nearly 100\% targets sensing accuracy across the entire SNR range, except for SNR $\leq$ –15 dB. In terms of overall performance, the three methods rank as follows: improved 1D-CNN $\ge$ enhanced DNN $\ge$ improved EDC.

  \item To further enable high-precision DOA estimation following source number sensing, a lightweight and adaptive DOA estimation method-online micro-clustering DOA(OMC-DOA) is proposed. By dynamically clustering angular candidates and updating direction estimates in real time, the proposed method achieves high estimation accuracy while significantly reducing computational complexity. Meanwhile, a closed-form expression of the Cramér–Rao lower bound (CRLB) for multi-source DOA estimation under the H²AD architecture is derived, serving as a theoretical performance benchmark. Furthermore, the steering vectors corresponding to different directions become asymptotically orthogonal in an ultra-massive antenna system (i.e., as $N_q \to \infty$). Simulation results demonstrate that the introduced OMC-DOA can approach the CRLB in medium-to-high SNR regimes.

\end{enumerate}

The remainder of this paper is organized as follows. Sec. \ref{sec:2} introduces the system model under the $ \text{H$\mathrm{^2}$AD} $ architecture. Then, three source number detection methods are proposed in Sec. \ref{sec:3}. In Sec. \ref{sec:4}, DOA estimation and theoretical analysis are presented. In Sec. \ref{sec:5}, we present the simulation results, followed by conclusion Sec. \ref{sec:6}.

Notations: $a$, $\mathbf{a}$, $\mathbf{A}$ denote scalars, vectors, matrices, respectively. $(\cdot)^{-1}$, $(\cdot)^T$ and $(\cdot)^H$ represent inverse, transposition, and conjugate transposition, respectively. $\text{diag}$ represents the diagonalization operation. $ \mathbb{E}\left[\cdot\right] $ denotes expectation. $\bigcup$ represents the union operation. $|\cdot|$ and $\text{std}(\cdot)$ represent the absolute value and the standard deviation, respectively. $\mathbf{Tr}(\cdot)$ denotes the matrix trace. $\otimes$ denotes the Kronecker product. $\dot{\mathbf{a}}$ is the partial derivative operation. $\mathbf{I}_M$ represents the identity matrix of size $M\times M$.

\section{System Model}\label{sec:2}


\begin{figure}[htp]
	\centering
    \includegraphics[width=0.45\textwidth]{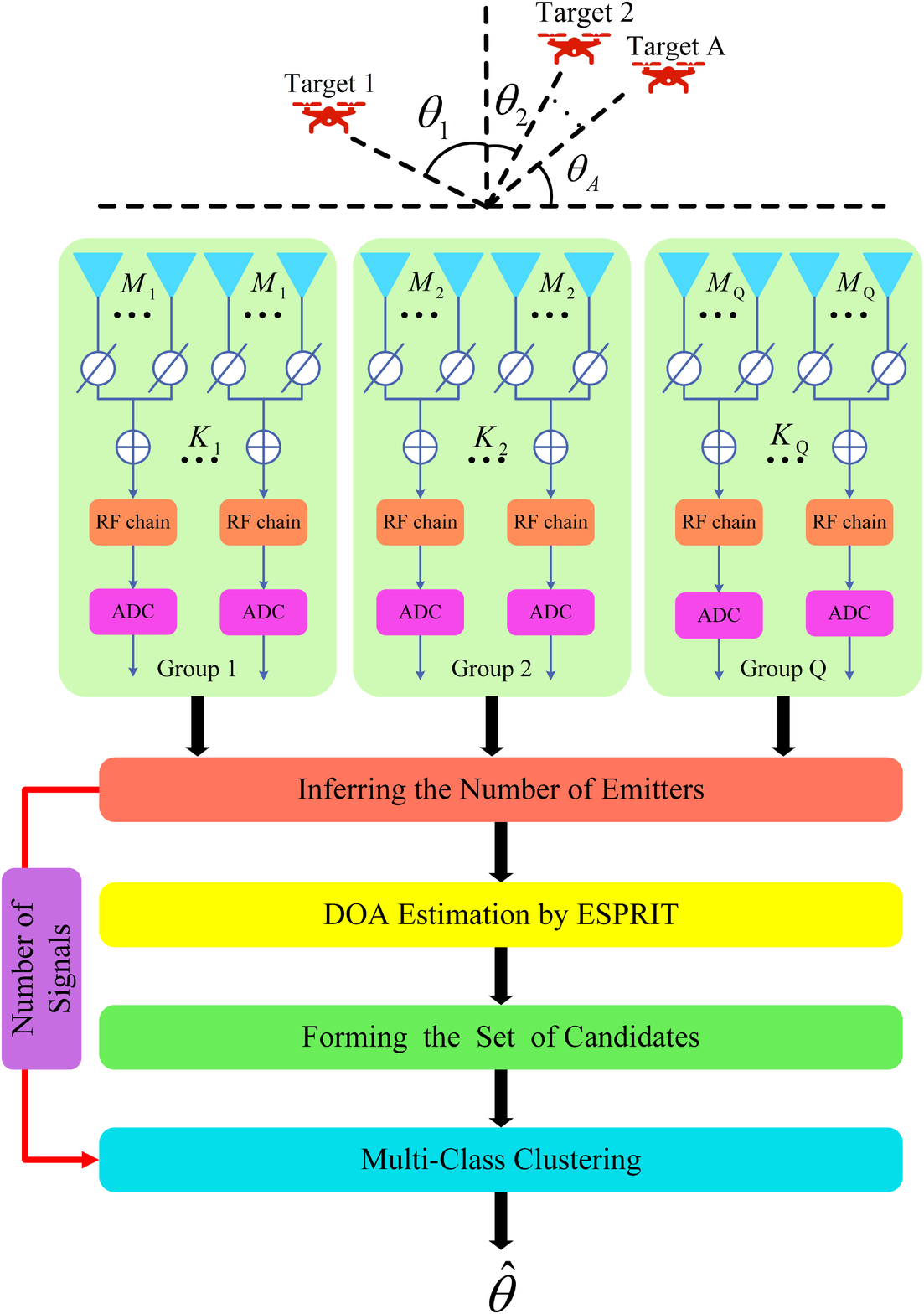}\\
	\caption{The framework of jointly sensing the number and direction values of multiple targets based on $ \text{H$\mathrm{^2}$AD} $ architecture.
    }\label{Fig-1}
\end{figure}



Fig.~\ref{Fig-1} illustrates a two-stage sensing framework that jointly estimates the number and direction values of multiple targets based on $ \text{H$\mathrm{^2}$AD} $ architecture. This framework consists of two stages: (1) target number sensing: the number of signal sources is estimated by extracting the eigenvalue information from the received signals, using approaches such as density-based clustering and data-driven neural networks. (2) multi-target DOA estimation: the ESPRIT algorithm is employed to generate the candidate DOAs set for multiple signal sources. Based on these candidates, a clustering and fusion-based method is applied to refine and integrate the results, thereby obtaining the final multi-source DOA estimations.
Define the set of target angles, the set of array group indices, the set of target indices, the set of antenna indices within each group, the set of antenna indices within each subarray, and the set of snapshots as $\varTheta=\{\theta_1, \theta_2, \dots, \theta_A\}$, \(\mathcal{G}=\{1, 2, \dots, Q\}\), \(\mathcal{T}=\{1, 2, \dots, A\}\), \(\mathcal{N}_q=\{1, 2, \dots, N_q\}\), \(\mathcal{M}_q=\{1, 2, \dots, M_q\}\), and \(\mathcal{T}_s=\{1, 2, \dots, T_s\}\), respectively.

Multiple targets from different directions are represented by their directions of arrival (DOAs), denoted as $\theta_i$ for $i \in\mathcal{T}$. 
Assuming $A$ far-field narrowband signal targets impinge on an H$2$AD MIMO antenna array, which is based on a uniform linear array (ULA) consisting of $M$ antennas partitioned into $Q$ groups. Each group contains $K_q$ subarrays, and $N_q$ antennas, with each subarray composed of $M_q$ antennas, satisfying ${K_1} = {K_2} = \cdots = {K_Q}$ and ${M_1} \ne {M_2} \ne \cdots \ne {M_Q}$.

For the $k$-th subarray of the $q$-th group, the signal received by the $m$-th antenna can be given by
\begin{equation}
\begin{aligned}
x_{q,k,m}(t) = \sum_{i=1}^{A}s_i(t) e^{j 2\pi (f_c t - f_c \tau_{q,k,m,i})} + w_{q,k,m}(t), \\
\end{aligned}
\end{equation}
where
$t (t \in [1,T])$ is the number of snapshots. $s(t)$ and $f_c$ denote the baseband signal and the carrier frequency, respectively. $w_{q,k,m}(t) \sim \mathcal{CN}(0, \sigma_w^2)$  represents noise vector. $\tau_{q,k,m,i}$ denotes the received $i$-th signal delay corresponding to the $m$-th antenna of the $k$-th subarray in the $q$-th group, which can be denoted as
\begin{equation}
\tau_{q,k,m,i} = \tau_0 - \frac{(km-1)d \sin (\theta_i)}{c}.
\end{equation}
where $\tau_0$ represents the  propagation delay from the incident signal to the receiver corresponding to reference antenna, $d$ is the distance between adjacent antennas, $\theta_i$ denotes the DOA of the $i$-th target, and  $c$ is the speed of light. Then, the received signal corresponding to the $k$-th subarray of the $q$-th group can be denoted as
\begin{equation}
y_{q,k}(t) = \frac{1}{\sqrt{M_q}}\sum_{m=1}^{M_q} x_{q,k,m}(t) e^{-j \phi_{q,k,m}},
\end{equation}
where $\phi_{q,k,m}$ stands for the adjustable phase corresponding to the $m$-th antenna. Thus, the received signal of $q$-th group can be expressed as
\begin{equation}
\mathbf{y}_q(t) = \mathbf{B}^{H}_{A,q} \mathbf{A}_q\mathbf{s}(t) + \mathbf{w}_q(t),
\end{equation}
where $
\mathbf{w}(t) =
\begin{bmatrix}
w_1(t), w_2(t), \cdots, w_{N_q}(t)
\end{bmatrix}^T\in \mathbb{C}^{N_q \times 1}$ denotes noise vector, satisfying $w_{n_q}\sim \mathcal{CN}(0,\sigma_{q}^2)$, $n_q\in \mathcal{N}_q$. $ \mathbf{B}_{A,q}\in \mathbb{C}^{M_q\times N_q}$ stands for a block matrix, where each block $\mathbf{b}_{A,q,k}$ representing the beamforming vector of the corresponding subarray,  given by 
\begin{equation}
\mathbf{b}_{A,q,k} =
\text{diag}(e^{j \phi_{q,k,1}} ,e^{j \phi_{q,k,2}} ,\cdots ,e^{j \phi_{q,k,M_q}}).
\end{equation}
And $\mathbf{A}_q\triangleq\mathbf{A}_q (\theta_1, \theta_2, \dots, \theta_A)\in \mathbb{C}^{M_q\times A}$ represents the array manifold matrix, denoted as
\begin{equation}
\mathbf{A}_q  =
\begin{bmatrix}
\mathbf{a}_q (\theta_1), \mathbf{a}_q (\theta_2), \dots, \mathbf{a}_q (\theta_A)
\end{bmatrix}.
\end{equation}
where $\mathbf{a}_q (\theta_i)$ denotes the array manifold vector of the signal impinging on the $q$-th subarray from direction $\theta_i$:
\begin{equation}
\mathbf{a}_q (\theta_i) =
\begin{bmatrix}
1,
e^{j \frac{2\pi}{\lambda} d \sin(\theta_i)},
\cdots,
e^{j \frac{2\pi}{\lambda} (N_q - 1) d \sin(\theta_i)}
\end{bmatrix}^{T}.
\end{equation}
The vector $\mathbf{s}(t)\in \mathbb{C}^{A \times 1}$ is the signal corresponding to $A$ independent sources:
\begin{equation}
\mathbf{s}(t) = \begin{bmatrix}
s_1(t), s_2(t), \cdots, s_A(t)
\end{bmatrix}^{T}.
\end{equation}
Through ADC, the signal can be converted into: 
\begin{equation}\label{10}
\mathbf{Y}_q = \mathbf{B}_{A,q}^H \mathbf{A}_q\mathbf{S} + \mathbf{W}_q,
\end{equation}
 $\mathbf{Y}_q=[\mathbf{y}_q(1), \mathbf{y}_q(2), \ldots, \mathbf{y}_q(T_s)]\in \mathbb{C}^{N_q\times T_s}$, $\mathbf{y}_q(t_s)$ denotes the received signal at the $t_s$-th ($t_s \in \mathcal{T}_s$) snapshot. $\mathbf{S}\in \mathbb{C}^{A\times T_s}$ and $\mathbf{W}_q\in \mathbb{C}^{N_q\times T_s}$ denote the source  signal matrix and the noise matrix, respectively.

To further estimate the number of signal sources, a covariance analysis of the received signal is performed.
According to Eq.~\eqref{10}, the corresponding covariance matrix can be given by
\begin{equation}
\mathbf{R}_q = \mathbb{E}\left[\mathbf{y}_q \mathbf{y}_q^H\right] = \sigma_s^2\mathbf{B}_{A,q} \mathbf{A}_q  \mathbf{A}_q^H \mathbf{B}_{A,q}^H + \sigma_v^2 \mathbf{I}_{N_q},
\end{equation}
where $\sigma_s^2$ and $\sigma_v^2$ denote the signal power and noise power, respectively. The covariance matrix contains the structure of both the signal subspace and the noise subspace. 
When the array dimension satisfies $N_q \ge A$, the covariance matrix can be expressed via eigenvalue decomposition as
\begin{equation}
\mathbf{R}_q = \mathbf{U}_q \mathbf{\Lambda}_q \mathbf{U}_q^H,
\end{equation}
where
$\quad\mathbf{\Lambda}_q = \text{diag}(\lambda_{q,1}, \lambda_{q,2}, \dots, \lambda_{q,N_q})$
is the diagonal matrix of eigenvalues arranged in descending order. Ideally, the first $A$ eigenvalues are significantly larger than the remaining ones, while the subsequent eigenvalues tend to converge to  the noise power $\sigma_v^2$. This distribution of eigenvalues reflects the distinction between the signal subspace and the noise subspace, thereby providing a theoretical basis for sensing the number of signal sources.

Based on the above analysis, the model of multi-targets angle values estimation can be introduced as follows.
Assume that the analog beamforming diagonal matrix is $\mathbf{b}_{A,q,k} =\text{diag}(\mathbf{1}_{1 \times M_q})$ and the received signal over 
$T$ snapshots can be expressed as
\begin{equation}
\mathbf{y}_{q,k}(t_s) = \mathbf{A}_{M_q}(\boldsymbol{\theta}) \mathbf{G}_q (\boldsymbol{\theta}) \mathbf{S}(t_s) + \mathbf{W}_q(t_s),
\end{equation}
where $\boldsymbol{\theta} \in \{\theta_1, \theta_2, \dots, \theta_A\}$, $\mathbf{S}(t_s) = \begin{bmatrix}
s_1(t_s), s_2(t_s), \cdots, s_A(t_s)
\end{bmatrix}^{T}$ is the signal vector.
The steering matrix $\mathbf{A}_{M_q}$ is denoted as:
\begin{equation}
\mathbf{A}_{M_q}(\theta_1, \dots, \theta_A) = 
\begin{bmatrix}
\mathbf{a}_{M_q}(\theta_1), \cdots, \mathbf{a}_{M_q}(\theta_A)
\end{bmatrix}.
\end{equation}
where $\mathbf{a}_{M_q}$ is expressed as:
\begin{equation}
\mathbf{a}_{M_q}(\theta_i) =
\begin{bmatrix}
1, e^{j \frac{2\pi}{\lambda} M_q d \sin \theta_i}, \cdots, e^{j \frac{2\pi}{\lambda} (K_q - 1) M_q d \sin \theta_i}
\end{bmatrix}^{T},
\end{equation}
And $\mathbf{G}_q (\boldsymbol{\theta})$ is the gain matrix, given by:
\begin{equation}
\mathbf{G}_q (\theta_1, \dots, \theta_A) =
\text{diag} \Big( {g}_q (\theta_1), \dots, {g}_q (\theta_A) \Big),
\end{equation}
where
\begin{equation}
\begin{aligned}
g_q(\theta_i) 
= \frac{1 - e^{j\frac{2\pi}{\lambda}M_qd\sin\theta_i}}
       {1 - e^{j\frac{2\pi}{\lambda}d\sin\theta_i}}.
\end{aligned}
\end{equation}
The ESPRIT algorithm is then applied to estimate the DOAs. Since each subarray can be modeled as an equivalent virtual array with an inter-element spacing of $M_q d$, the estimation results inherently have $2\pi$ phase ambiguity. 
Each subarray group yields multiple possible candidate DOAs for each signal, with the number of candidates determined by the number of antennas $M_q$ in the corresponding subarray. The candidate angles set across all subarray groups and signals can be collectively represented as

\begin{equation}
\hat{\Theta} = \left\{ \hat{\theta}_{q,i,j} \,\middle|\, q \in \mathcal{G};\; i \in \mathcal{T};\; j \in \mathcal{M}_q \right\},
\end{equation}
The total number of candidate DOAs is given by:
\begin{equation}
\left| \hat{\Theta} \right| = A \sum_{q=1}^Q M_q.
\end{equation}

In this set, due to noise-induced perturbations, the true DOAs $\theta_i$ estimated from different subarrays tend to cluster tightly around the actual direction values. In contrast,  the pseudo-solutions are typically scattered across the angular domain:
\begin{equation}
\theta_i + \Delta_{q,m}.
\end{equation}
where $\Delta_{q,m} = \frac{2\pi m}{M_q}, m = \{1, \dots, M_q - 1\}$. Notably, when all $M_q$ are pairwise coprime, the pseudo-offsets among different subarrays do not overlap, , preventing the formation of false clusters in the candidate set $\hat{\Theta}$. Therefore, the final step is to identify $A$
true DOA clusters from the candidate set
$\hat{\Theta}$.




\section{Proposed Source Number Detection Methods}\label{sec:3}
In multi-source sensing tasks, accurately estimating the number of signal sources forms the foundation for subsequent processing steps. This task becomes particularly challenging in scenarios with low SNR or severe interference. To enhance both accuracy and robustness, this section proposes three source number estimation methods: an improved EDC, an enhanced DNN, and an improved 1D-CNN.

\subsection{Improved EDC Method}

\begin{figure}[htp]
	\centering
    \includegraphics[width=0.30\textwidth]{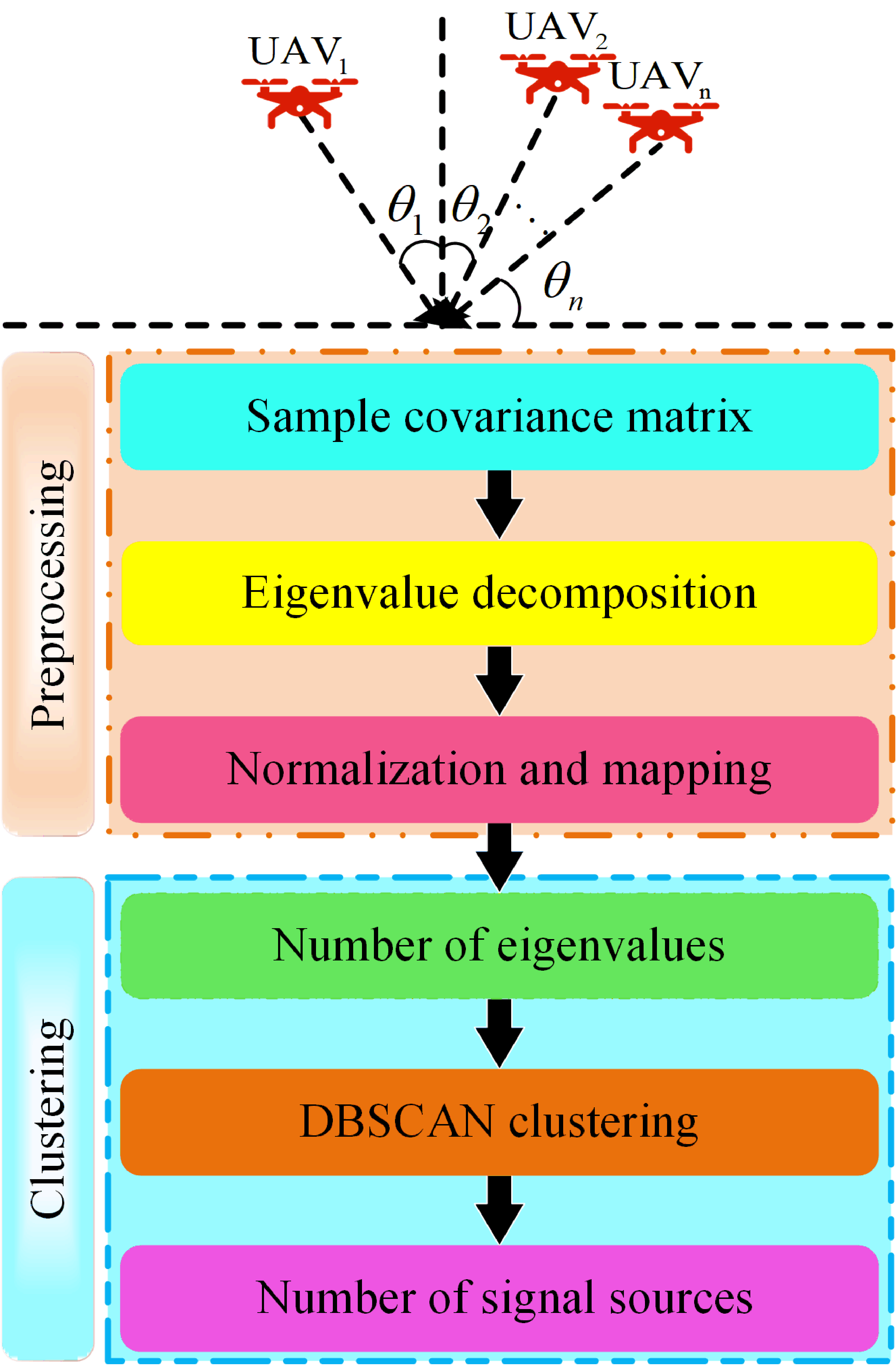}\\
	\caption{ Illustration of the improved EDC method.}\label{Fig-5}
\end{figure}

As the number of antennas increases, reception performance can be enhanced in practical signal processing,
thereby amplifying the disparity between signal and noise subspaces. This manifests as larger eigenvalues corresponding to the signal subspace compared to those of the noise subspace when performing eigenvalue decomposition of the signal covariance matrix. Therefore, the hybrid heterogeneous array has a good ability to distinguish the number of information sources. As illustrated in Fig.~\ref{Fig-5}, the proposed improved EDC method primarily consists of two steps: preprocessing and clustering.  Specifically, in the preprocessing, eigendecomposition is used to obtain all the eigenvalues of the covariance matrix corresponding to received signal.  Then, clustering is carried out after standardization and mapping operations. The normalized eigenvalue $z_{q,j}$ corresponding to the $q$-th group can be expressed as

\begin{equation}
z_{q,j} = \frac{\lambda_{q,j} - \mu_q}{\sigma_q},
\end{equation}
where
\begin{equation}
\mu_q = \frac{1}{N_q}\sum_{j=1}^{N_q} \lambda_{q,j},
\end{equation}
\begin{equation}
\sigma_q = \sqrt{\frac{1}{N_q}\sum_{j=1}^{N_q} (\lambda_{q,j} - \mu_q)^2} .
\end{equation}
Here, $j$ represents the index related to the dimension of the covariance matrix. To further enhance the estimation accuracy of the number of sources, the one-dimensional eigenvalues $z_{q,j}$ $(\forall q,j)$ are mapped to the two-dimensional coordinates $(x_{q,j},\ y_{q,j})$ as follows
\begin{equation}\label{23}
(x_{q,j},\ y_{q,j}) = (z_{q,j},\ z_{q,j}^{\epsilon}),
\end{equation}
where $\epsilon$ denotes higher-order index. In \eqref{23}, $x_{q,j}$ and $y_{q,j}$ represent the lower and higher orders of $z_{q,j}$, respectively. The mapping introduces additional dimensions corresponding to higher-order terms, thereby enhancing data separability.   Thus, the eigenvalues associated with signals become more distinct and readily separable from the cluster of noise eigenvalues. Correspondingly, the set of eigenvalues corresponding to the $q$-th group can be represented as
\begin{equation}
\mathcal{Z}_q = \left\{ (x_{q,i}, y_{q,i}) \right\}_{i=1}^{N_q}.
\end{equation}
Then, the set of eigenvalues from $Q$ groups can be given by
\begin{equation}
\mathcal{Z} = \bigcup_{q=1}^{Q} \mathcal{Z}_q.
\end{equation}
Thus, for all groups, the difference in the eigenvalues corresponding to the signal and noise becomes larger. Specifically, the eigenvalues related to noise usually have lower values on different subarrays. After standardization and mapping, eigenvalues of the noise are concentrated near the origin in a two-dimensional space, forming a compact high-density cluster. In contrast, the eigenvalues related to signals usually have a large values, and their distributions are further away from the origin, forming multiple sparse local clusters. Thus, signal sources and noise can be effectively distinguished based on the eigenvalues. Further, in the global feature space $\mathcal{Z}$, the density-based noisy application space clustering (DBSCAN) algorithm is adopted for unsupervised clustering and to estimate the number of signal sources.

Let the main dense cluster identified by DBSCAN be denoted as $\mathcal{C}_{\text{noise}}$, which corresponds to the large number of densely distributed noise feature points in the space. Its complement $\mathcal{C}_{\text{signal}} = \mathcal{Z} \setminus \mathcal{C}_{\text{noise}}$ is regarded as the sparse cluster set associated with signal features. Accordingly, the final estimated number of targets is obtained by normalizing the number of point targets in the signal-related cluster by the number of array groups, which is given by
\begin{equation}
\hat{A} = \frac{|\mathcal{C}_{\text{signal}}|}{Q}.
\end{equation}
Then, the number of targets can be obtained.

\subsection{Enhanced DNN}

\begin{figure}[htp]
	\centering
    \includegraphics[width=0.45\textwidth]{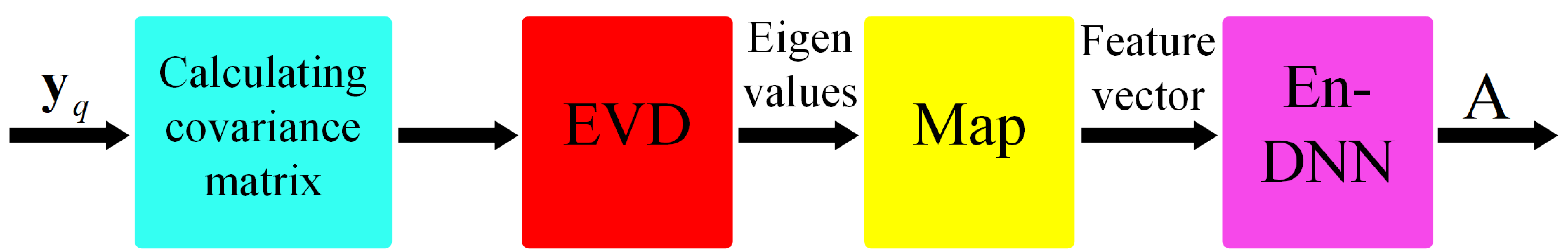}\\
	\caption{Illustration of the enhanced DNN method.}\label{Fig-3}
\end{figure}

\begin{figure}[htp]
	\centering
    \includegraphics[width=0.45\textwidth]{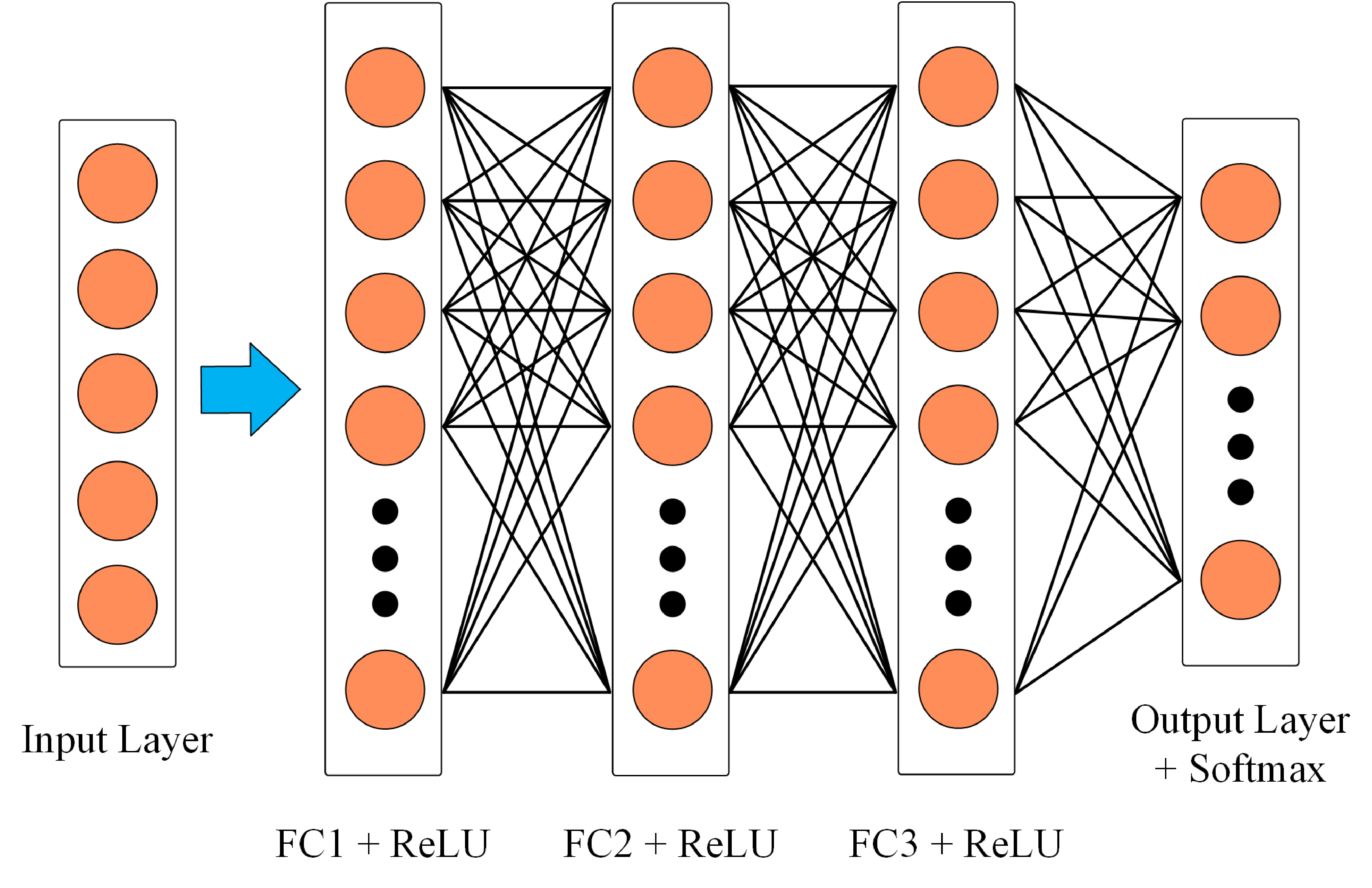}\\
	\caption{Network architecture of the enhanced DNN.}\label{Fig-7}
\end{figure}

To estimate the number of signal sources under low SNR conditions, a supervised learning method based on an enhanced DNN is proposed in this section. 
Fig.~\ref{Fig-3} illustrates the method framework of enhanced DNN, which predicts the number of sources $A$ using a trained neural network based on the statistical features of eigenvalues extracted from the sample covariance matrix of the received signal $\mathbf{y}_q$. The procedure includes covariance matrix computation, eigenvalue decomposition, statistical feature extraction, and deep neural network inference.
Specifically, for each observation, $M$ eigenvalues $\{ \lambda_i \}_{i=1}^{M}$ are extracted from the covariance matrix, where $M$ is the total number of antennas for $\text{H$\mathrm{^2}$AD}$. Subsequently, five statistical features are extracted from the eigenvalues, which can be used to characterize the number of signal sources and provide potential basis for the sensing for the number of targets. Then, logarithmic transformation is performed on these statistical features to compress their dynamic range, thereby enhancing the discrimination under different numbers of signal sources. The extracted features can be denoted as
\begin{equation} \left\{
\begin{aligned}
&\beta_1 = f_1(\lambda_1,\lambda_2,\cdots,\lambda_{M}) ,\\
&\beta_2 = f_2(\lambda_1,\lambda_2,\cdots,\lambda_{M}), \\
&\beta_3 = f_3(\lambda_1,\lambda_2,\cdots,\lambda_{M}), \\
&\beta_4 = f_4(\lambda_1,\lambda_2,\cdots,\lambda_{M}), \\
&\beta_5 = f_5(\lambda_1,\lambda_2,\cdots,\lambda_{M}).
\end{aligned}
\right.
\end{equation}

The mapping of the characteristic statistics of all parameters can be written as
\begin{equation} \left\{
\begin{aligned}
\beta_1 &= \log\left( \max \lambda_i \right),\\
\beta_2 & = \log\left( \min \lambda_i \right),\\
\beta_3 &= \log\left( \text{std}(\{\lambda_i\}) \right), \\
\beta_4 &= \log\left( \frac{1}{M} \sum_{i=1}^{M} \lambda_i \right), \\
\beta_5 &= -\sum_{i=1}^{M} \left( \frac{\lambda_i}{\sum_{j=1}^{M} \lambda_j} \right) \log\left( \frac{\lambda_i}{\sum_{j=1}^{M} \lambda_j} \right),
\end{aligned}\right.
\end{equation}
where $\beta_1$ and $\beta_2$ represent the maximum and minimum eigenvalues, characterizing the boundary properties of the eigenvalue spectrum. $\beta_3$ and $\beta_4$ denote the standard deviation and mean value, respectively, quantifying the overall dispersion and energy distribution characteristics. $\beta_5$ corresponds to the eigenvalue entropy, which measures the energy concentration degree. As the number of signal sources increases, the entropy exhibits a monotonic increasing trend, thereby providing sensitive detection of source variations and enhanced discrimination accuracy. The set of five statistical features can be represented as
\begin{equation}\label{29}
\mathbf{x} = [\beta_1, \beta_2, \beta_3, \beta_4, \beta_5]^T.
\end{equation}
Thus, the eigenvalue data are compressed into a low-dimensional feature vector while retaining the essential statistical information required for distinguishing between signal and noise subspaces. 

Based on \eqref{29},  an enhanced DNN is designed, as illustrated in Fig.~\ref{Fig-7}. This network consists of three fully connected layers, each of which is equipped with ReLU activation and Dropout layers to enhance nonlinear expression capabilities and alleviate overfitting.
The output layer uses a Softmax activation function to produce a probability distribution over the possible number of sources, thereby enabling classification-based prediction. The input $\mathbf{h}^{(\ell-1)}$ corresponding to the $\ell$-th layer can be denoted as
\begin{equation}
	\mathbf{h}^{(\ell)} = \mathrm{ReLU}\left(W^{(\ell)} \mathbf{h}^{(\ell-1)} + \mathbf{b}^{(\ell)}\right), \ \ell = 1, 2, \ldots, \hat{L},
\end{equation}
where $\hat{L}=3$ denotes the number of fully connected layers. Then, the forward propagation of the proposed neural network can be given by
\begin{equation}
\hat{\mathbf{y}} = \mathrm{Softmax}\left(W^{(4)} \mathbf{h}^{(3)} + \mathbf{b}^{(4)}\right),
\end{equation}
where $W ^ {4} $  and $b ^ {4} $ denote the weight matrix and bias vector of $4 $-th layer, respectively. ReLU is a linear rectification activation function used to introduce nonlinearity, while the Softmax function maps the output to a probability distribution. The ReLU function is used as a nonlinear activation to enhance the representational ability of the network, while the Softmax function converts the logits input at the output layer into a probability distribution of the number of signal sources.
$\hat{\mathbf{y}} = [\hat{y}_1, \hat{y}_2, \dots, \hat{y}_A]^T$ denotes the predicted probability distribution over the $A$ candidate source numbers, where $\hat{y}_i$ ($i\in\mathcal{T}$) represents the predicted probability corresponding to $i$ sources. The final estimate is given by the class corresponding to the maximum probability.
During training, the network parameters are optimized using the cross-entropy loss function, which is repressed as
\begin{equation}
\label{32}
\mathcal{L} = -\sum_{c=1}^{A} y_a \log \hat{y}_a,
\end{equation}
where $y = [y_1, y_2, \dots, y_A]^T$ is the one-hot encoded ground-truth label vector. The parameters $\{W^{(\ell)}, b^{(\ell)}\}$ are iteratively updated via backpropagation in conjunction with gradient descent, yielding a discriminative model with improved generalization capability.


\subsection{Improved 1D-CNN}

\begin{figure}[htp]
	\centering
    \includegraphics[width=0.40\textwidth]{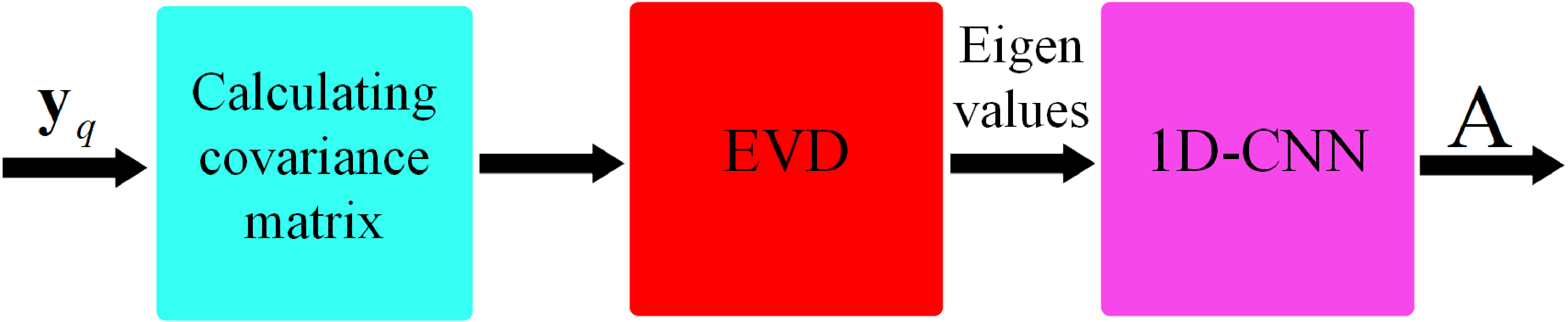}\\
	\caption{Illustration of the improved 1D-CNN-based method.}\label{Fig-4}
\end{figure}

\begin{figure}[htp]
	\centering
    \includegraphics[width=0.49\textwidth]{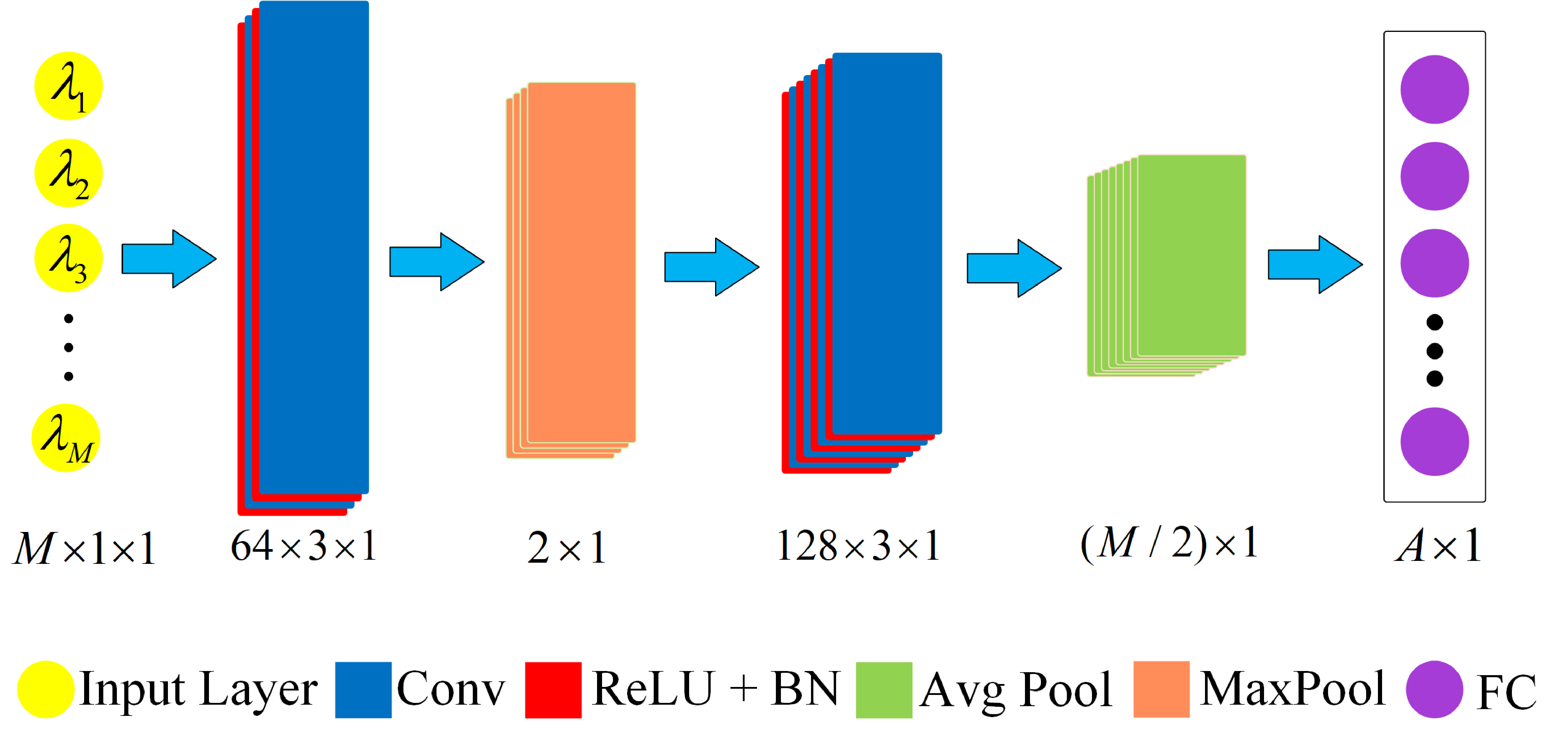}\\
	\caption{The architecture of the improved 1D-CNN.}\label{Fig-6}
\end{figure}

Considering that manually constructed statistical features may struggle to comprehensively capture the variation patterns of the number of signal sources, this section proposes an improved 1D-CNN method, as illustrated in Fig~\ref{Fig-4}. By directly learning source number-related patterns from the spectral sequence, this method improves both generalization and noise robustness.

Specifically, we exploit the eigenvalues $\lambda_1,\lambda_2,\cdots,\lambda_{M}$ obtained from the sample covariance matrix as the input. Since these eigenvalues typically show exponential attenuation at low SNRs, the magnitudes of the eigenvalues vary significantly under different numbers of sources. Directly feeding raw eigenvalues into the network can cause scale imbalance, increased training difficulty, and unstable gradient propagation. To mitigate this, a logarithmic transformation is applied to compress the dynamic range of the eigenvalues, enhancing the comparability of spectral structures under varying conditions and improving network training stability and classification capability. The transformed input $\mathbf{x}$ vector can be constructed as
\begin{equation}
\mathbf{x}=[\log{\lambda_1},\log{\lambda_2},\cdots,\log{\lambda_{M}}]^T\in R^{M\times 1}.
\end{equation}
Here, $\mathbf{x}$ is reshaped into a tensor of size $[M \times 1 \times 1]$ and fed into the improved 1D-CNN, facilitating automatic feature extraction. Fig.~\ref{Fig-6} illustrates the architecture of the improved 1D-CNN network. The network consists of two convolutional modules followed by a fully connected classifier, effectively capturing eigenvalues for source number estimation. The first convolutional layer employs 64 filters with a $3 \times 1$ kernel to capture local variations between adjacent eigenvalues and detect energy transitions caused by changes in the number of sources. Batch normalization and ReLU activation are then applied to enhance nonlinear representation, followed by a $2 \times 1$ max-pooling layer for downsampling, which reduces redundancy and improves robustness to scale variations. The second convolutional layer uses a $3 \times 1$ kernel and expands the number of channels to 128 to extract higher-order spectral features. Finally, ReLU activation and global average pooling (GAP) are applied to compress spatial dimensions and enhance structural generalization.


After feature extraction, the network output is fed into a fully connected layer with $A$ neurons, where $A$ denotes the maximum number of signal sources. The output is passed through a Softmax function to generate a probability prediction vector $\hat{\mathbf{y}}$. 
Under the supervised learning framework, the network uses the $\mathcal{L}$ to train defined in~\eqref{32}. The network directly extracts discriminative features from the eigenvalues, thereby achieving effective sensing of the number of targets.

\section{DOA Estimation and Theoretical Analysis}\label{sec:4}
\subsection{Multi-Targets DOA Estimation}
The above three develpoed approaches, improved EDC, enhanced DNN, and improved 1D-CNN have provided the critical prerequisite for multi-source DOA estimation, thereby effectively avoiding redundant computations and reducing the computational cost in DOA estimation. For clarity, the multi-targets DOA clustering and fusion-based method is introduced  in the following. 

Similar to single-source clustering methods, this paper applies the online micro-clustering DOA estimation (OMC-DOA) technique to multi-source scenarios, building upon the weighted global minimum distance (WGMD) and weighted local minimum distance (WLMD) frameworks introduced in~\cite{10767772}. The proposed method incorporates three key components: (1) dynamic micro-cluster updating, (2) an exponential decay mechanism, and (3) inter-cluster fusion strategies. It enables real-time selection of optimal DOA estimates from redundant candidate sets, deriving final source bearings ($A$-signal) through online processing of all subarray candidates. OMC-DOA demonstrates superior real-time performance and computational efficiency, significantly enhancing system flexibility and operational effectiveness.


\subsection{Derivation of CRLB}

To evaluate the theoretical performance limit of DOA estimation under the $ \text{H$\mathrm{^2}$AD} $ architecture, the closed-form expression of the Cramér–Rao lower bound (CRLB) is derived. Unlike traditional single-source scenarios under the $ \text{H$\mathrm{^2}$AD} $ architecture in~\cite{10767772}, the derived CRLB demonstrates unbiased estimation of multiple signal sources.

Specifically, the lower bound of the estimation variance satisfies
\begin{equation}
\label{eq:crlb}
\mathrm{Cov}(\hat{\boldsymbol{\theta}}) \geq \frac{1}{N}\, \mathbf{FIM}^{-1},
\end{equation}
where \(\hat{\boldsymbol{\theta}} = [ \hat{\theta}_1, \hat{\theta}_2, \dots, \hat{\theta}_A ]^T\) denotes the estimated DOAs of all signal sources.
The inverse of the Fisher information matrix (FIM) in~\eqref{eq:crlb} can be expressed as
\begin{equation}
\label{eq:fim}
\frac{1}{N}\, \mathbf{FIM}^{-1} =
\operatorname{diag}\left(\frac{1}{N\sum_{q=1}^Q\alpha_{q,1}}, \dots, \frac{1}{N\sum_{q=1}^Q\alpha_{q,n}}\right).
\end{equation}
where the closed-form expression of \(\alpha_{q,i}\) is given by~\eqref{eq:alpha}, as illustrated at the top of this page. The reciprocal of FIM is CRLB. The specific derivations are shown in Appendix \ref{appendices}.

\begin{figure*}[!t]
\begin{equation}
\label{eq:alpha}
\alpha_{q,i}
\!\!=\!\!
\Biggl(
L \sum_{q=1}^{Q}
\frac{8\pi^{2}\,\sigma^{4}\,\cos^{2}\theta_{i}
\Bigl[
\|\mathbf{g}_{q,i}\|^{4}\,\bigl(N_{q}\,\nu_{q,i} - \mu_{q,i}^{2}\bigr)\,\bigl(\gamma\,N_{q}\,\|\mathbf{g}_{q,i}\|^{2} + M_{a}\bigr)
\;+\;M_{a}\,N_{q}^{2}\,\bigl(\|\mathbf{g}_{q,i}\|^{2}\,\|\bm{\Gamma}_{i}\|^{2}
-\mathbb{R}[(\bm{\Gamma}_{i}^{H}\,\mathbf{g}_{q,i})^{2}\bigr]\bigr)
\Bigr]}
{\lambda^{2}\,M_{a}\,\bigl(\gamma\,N_{q}\,\|\mathbf{g}_{q,i}\|^{2} + M_{a}\bigr)^{2}}
\Biggr)^{-1}.
\end{equation}\hrulefill
\end{figure*}

\subsection{Computational Complexity}
\begin{table}[]
	\centering
	\caption{The caption of this table}\normalsize
	\begin{tabular}{l|c}
		\toprule
Algorithms & Complexity \\
\midrule
		FCNN\cite{WOS:000979292100001} & $
		2(LH+2H^2+HA)
		$ \\
		Proposed enhanced DNN & $
		2(LH+2H^2+HA)
		$ \\
		Improved 1D-CNN & $
		24960L+256A
		$ \\
		\bottomrule
	\end{tabular}
	\label{tab:complexity}
\end{table}
%

To further compare the computational complexity of developed approaches, a theoretical analysis of the forward propagation FLOPs is conducted for the two proposed neural network structures, as well as the fully connected neural network (FCNN) model introduced in~\cite{WOS:000979292100001}. Let $L$ be the input feature dimension, $H$ the number of neurons in the hidden layers, and $A$ the number of output classes. For the three-layer FCNN and the enhanced DNN with ReLU and Dropout, the  complexity can be computed as $2(LH + 2H^2 + HA)$ FLOPs. For the improved 1D-CNN model, assuming two convolutional layers with 32 and 64 filters, kernel size $A$, and the output layer with 256 neurons, the overall complexity can be approximated by $24960L + 256A$ FLOPs. The detailed comparison is presented in Table~\ref{tab:complexity}.


\section{Simulation Results}\label{sec:5}
In this section, the simulation results of the proposed methods are presented.  Specifically, we evaluate the three proposed methods for estimating the number of multi-source signals, respectively. Subsequently, to validate the effectiveness of the proposed framework, the DOA estimation performance is rigorously analyzed. 


For the estimation of the number of signal sources, the simulation configuration is as follows: the number of signal sources is $A=3$, the number of snapshots is 200, and the noise power is 30 dBm. All results are obtained from 5000 Monte Carlo trials. The total number of receiving antennas is $M = 97$. The eigenvalue clustering method divides them into three subarrays with sizes $N_q \in \{29, 31, 37\}$ for feature extraction, whereas the enhanced DNN model and the 1D-DNN model directly use the $M$-dimensional vector as input. The number of network layers and the size of the training dataset for the FCNN model have been aligned with those of the enhanced DNN model to ensure a fair comparison. For DOA estimation, the true DOAs are set as $\boldsymbol{\theta} = \{11^\circ,\ 23^\circ\}$. The receiver array comprises $Q = 3$ array groups, each containing $K = 16$ subarrays. The number of antennas per subarray is $M_1=7$, $M_1=13$, and $M_1=17$. Each trial uses 200 snapshots, and 5000 Monte Carlo runs are conducted to ensure statistically stable RMSE results.

\begin{figure*}[t]  
    \centering
    \begin{subfigure}[b]{0.32\textwidth}
        \includegraphics[width=\linewidth]{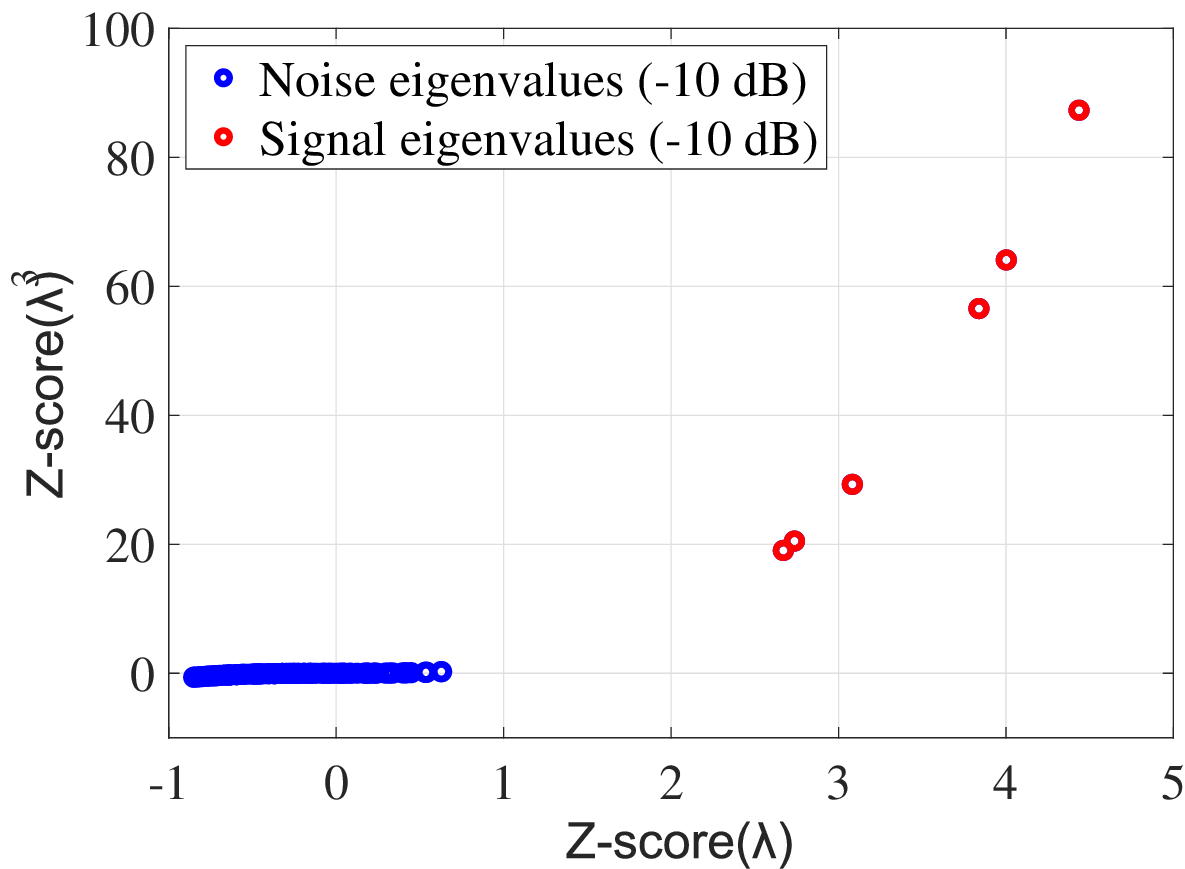}
        \caption{SNR = –10 dB}
        \label{fig:sub1}
    \end{subfigure}
    \hfill
    \begin{subfigure}[b]{0.32\textwidth}
        \includegraphics[width=\linewidth]{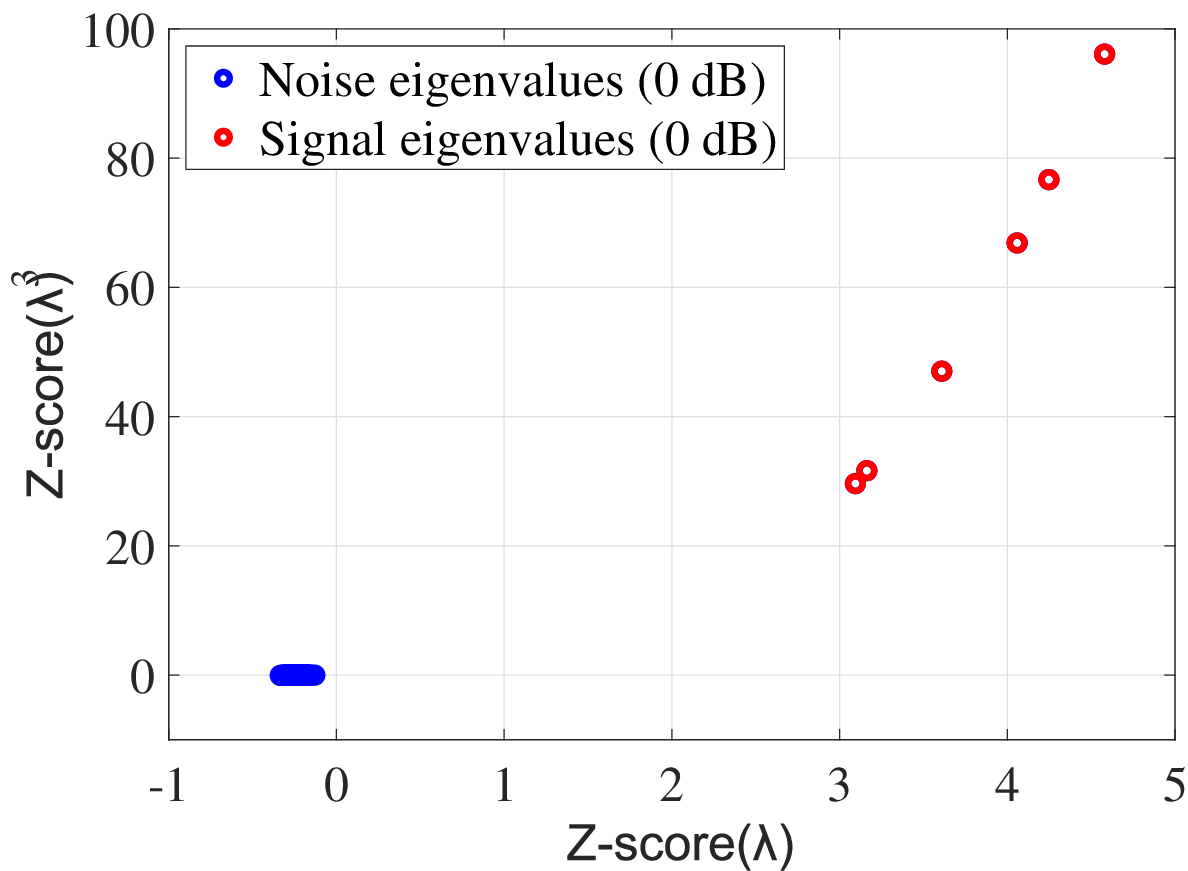}
        \caption{SNR = 0 dB}
        \label{fig:sub2}
    \end{subfigure}
    \hfill
    \begin{subfigure}[b]{0.32\textwidth}
        \includegraphics[width=\linewidth]{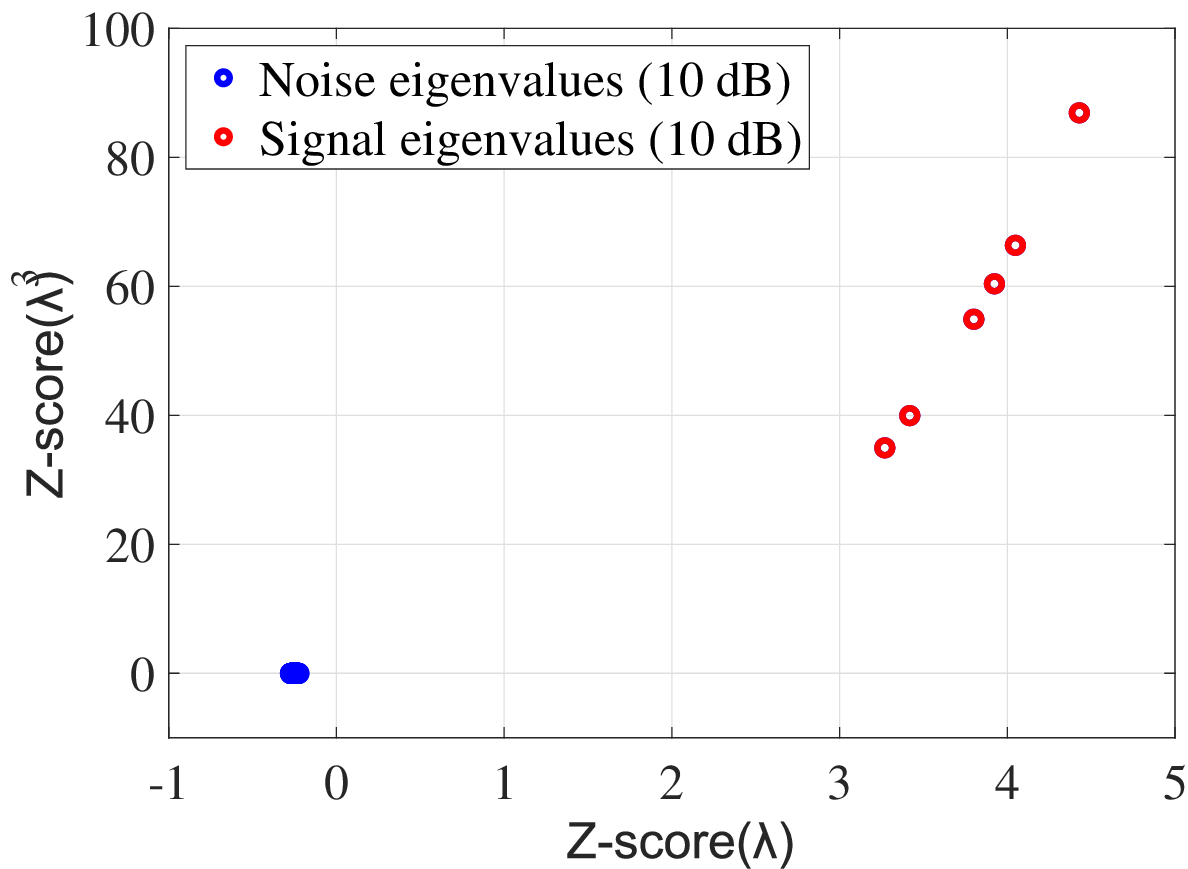}
        \caption{SNR = 10 dB}
        \label{fig:sub3}
    \end{subfigure}
    \caption{The distribution of eigenvalues under different SNRs.}
    \label{fig_scatterplot}
\end{figure*}

To intuitively illustrate the performance of the eigenvalue-domain clustering method under varying SNR conditions, Fig.~\ref{fig_scatterplot} presents the eigenvalue scatter distributions across different SNR levels. As shown, at SNR $=$ -10dB, the noise eigenvalues are highly concentrated and clearly separated from the signal-related eigenvalues, resulting in distinct clustering outcomes. Even under low-SNR scenarios such as SNR $=$ -10dB, despite the stronger influence of noise, the mapped feature points still exhibit visible clustering structure. This indicates that the method possesses inherent noise robustness due to its structural design. The sustained inter-class separability at low SNR is primarily attributed to the higher energy associated with the signal subspace eigenvalues, which tend to stand out during the feature mapping process. This geometric distinction between signal and noise clusters provides a favorable basis for subsequent clustering-based classification.

\begin{figure}[htp]
	\centering
    \includegraphics[width=0.48\textwidth]{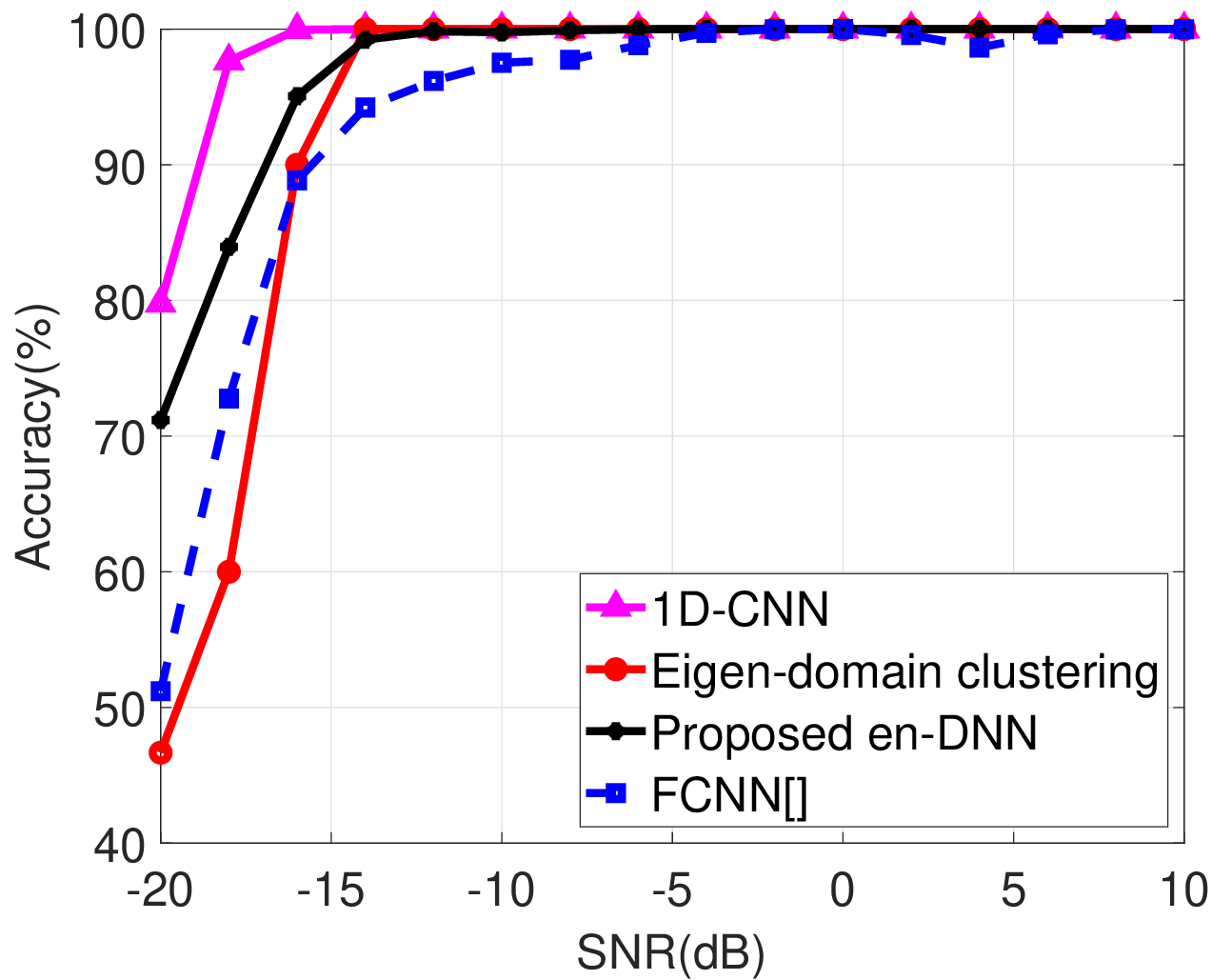}\\
	\caption{Estimation accuracy versus SNR for multiple source detection.}\label{Fig_signal}
\end{figure}

As shown in Fig.~\ref{Fig_signal}, the estimation accuracy of the number of signal sources by the three proposed methods and the reference scheme (FCNN) in~\cite{WOS:000979292100001} under different SNR conditions is compared. The results indicate that the three proposed methods generally outperform the traditional baseline model across all SNR levels, especially demonstrating higher accuracy in low-SNR scenarios. Among them, the improved 1D-CNN model achieves the best performance across the entire SNR range, with an accuracy of nearly 80\% even at -20 dB, significantly surpassing the other methods. The enhanced DNN model shows a steady improvement in accuracy under low SNR conditions and overall performs better than the eigenvalue domain clustering method and FCNN. The eigenvalue domain clustering method exhibits a rapid increase in accuracy and tends to saturate when SNR $\ge$ -14 dB, but is inferior to FCNN in the low-SNR region. Overall, the improved 1D-CNN model demonstrates the strongest adaptability in high-noise environments, the enhanced DNN model achieves a balance between complexity and performance, and the clustering method shows high accuracy in medium-to-high SNR conditions. 

\begin{figure}[t]
    \centering
    \begin{subfigure}{0.48\textwidth}
        \centering
        \includegraphics[width=\textwidth]{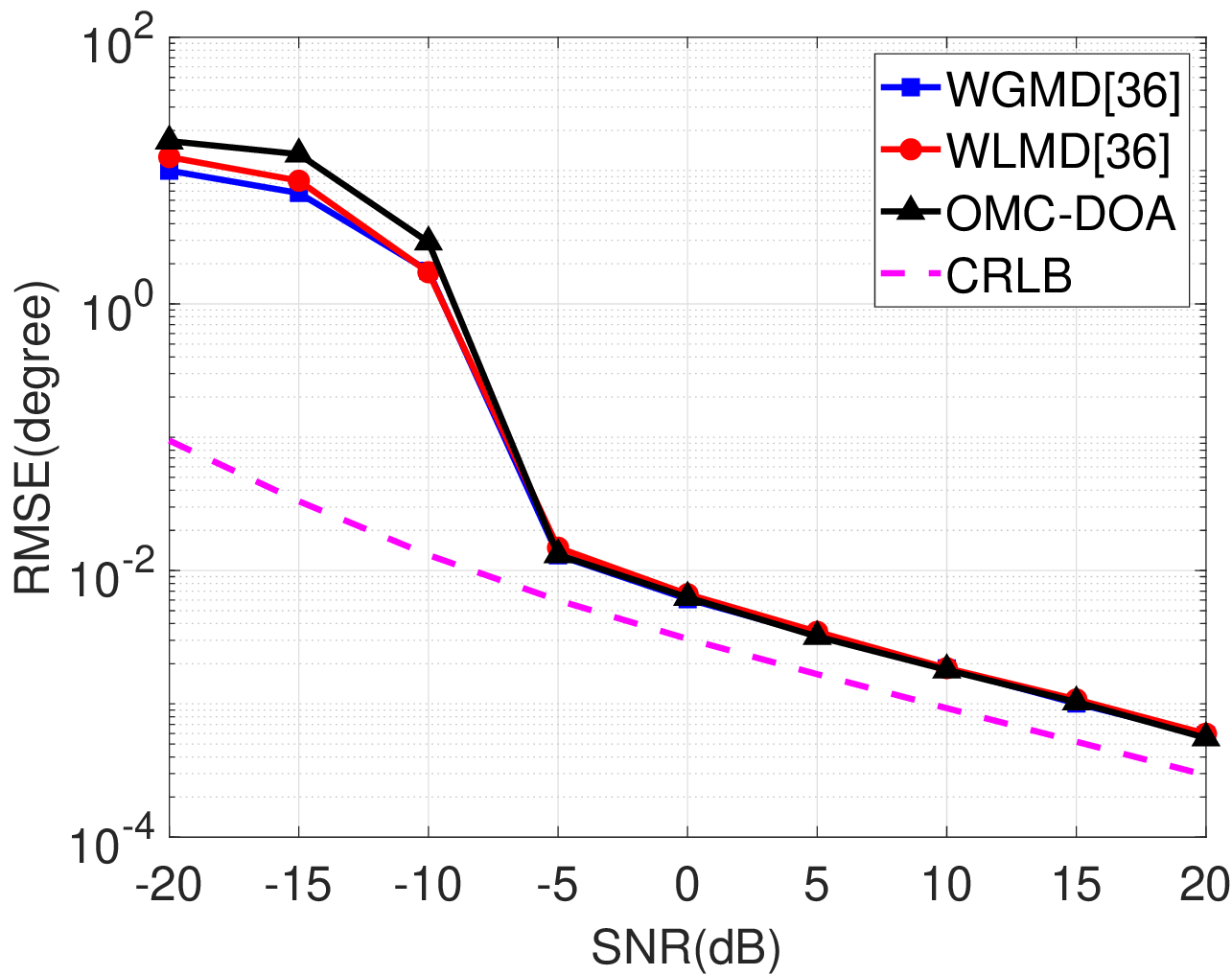}
        \caption{$\theta=11^{\circ}$}
        \label{Fig_emse1}
    \end{subfigure}
    \vspace{0.5em}  
    \begin{subfigure}{0.48\textwidth}
        \centering
        \includegraphics[width=\textwidth]{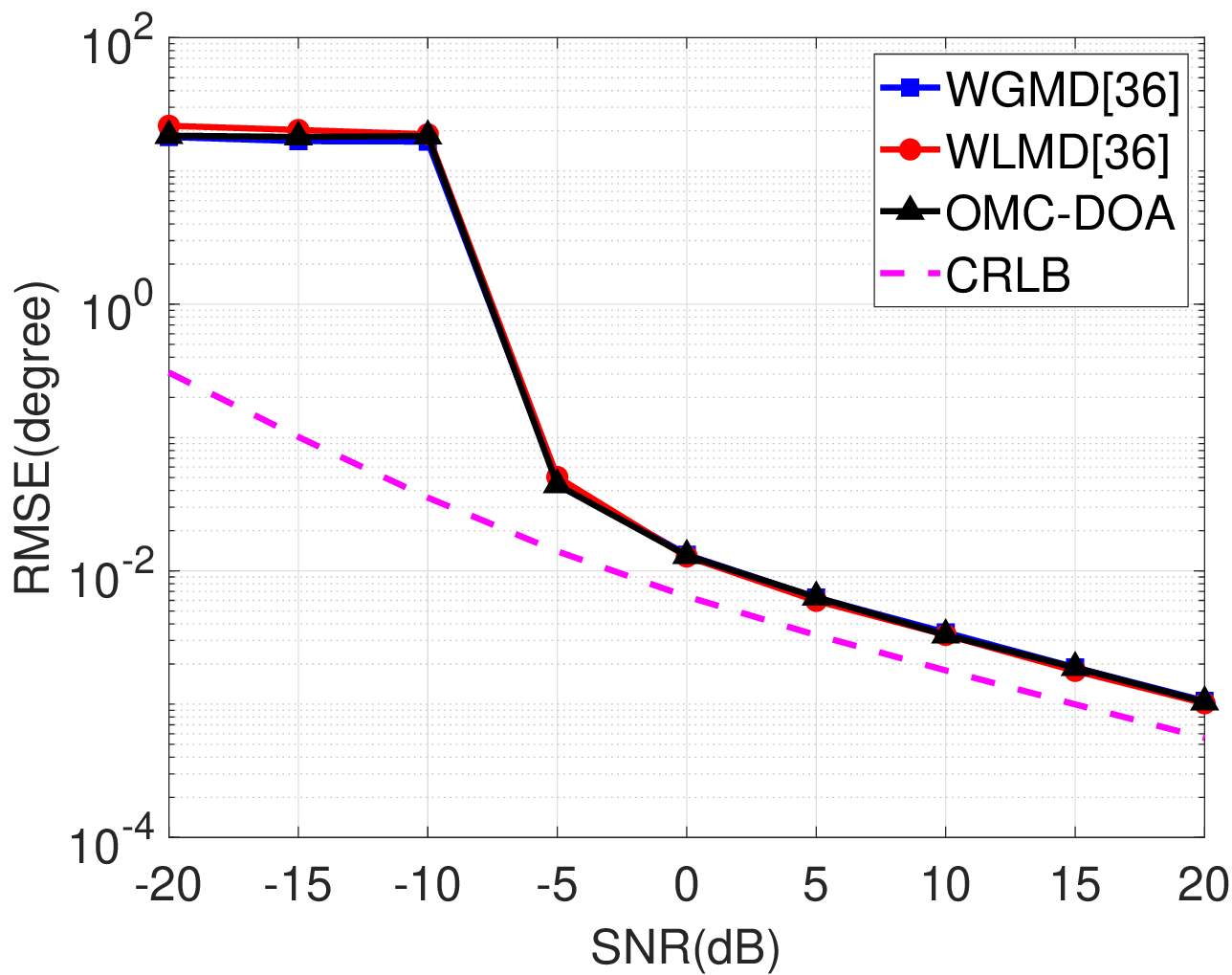}
        \caption{$\theta=23^{\circ}$}
        \label{Fig_rmse2}
    \end{subfigure} 
    \caption{RMSE versus SNR for DOA estimation.}
    \label{Fig_rmse_all}
\end{figure}

To evaluate the localization performance, we conduct comprehensive simulations for multi-source DOA estimation. The proposed OMC-DOA algorithm is compared with the WGMD and WLMD algorithms presented in~\cite{10767772}. Fig. \ref{Fig_rmse_all} presents the root-mean-square error (RMSE) performance comparison of the three clustering methods at two distinct angular positions across varying SNR regimes. The experimental results demonstrate that WGMD and WLMD achieve marginally superior performance compared to OMC-DOA within the SNR range of -20 dB to -5 dB. Notably, all three methods exhibit identical RMSE performance when SNR reaches -5 dB.

\begin{figure}[htp]
    \centering
    \begin{subfigure}{0.48\textwidth}
        \centering
        \includegraphics[width=\textwidth]{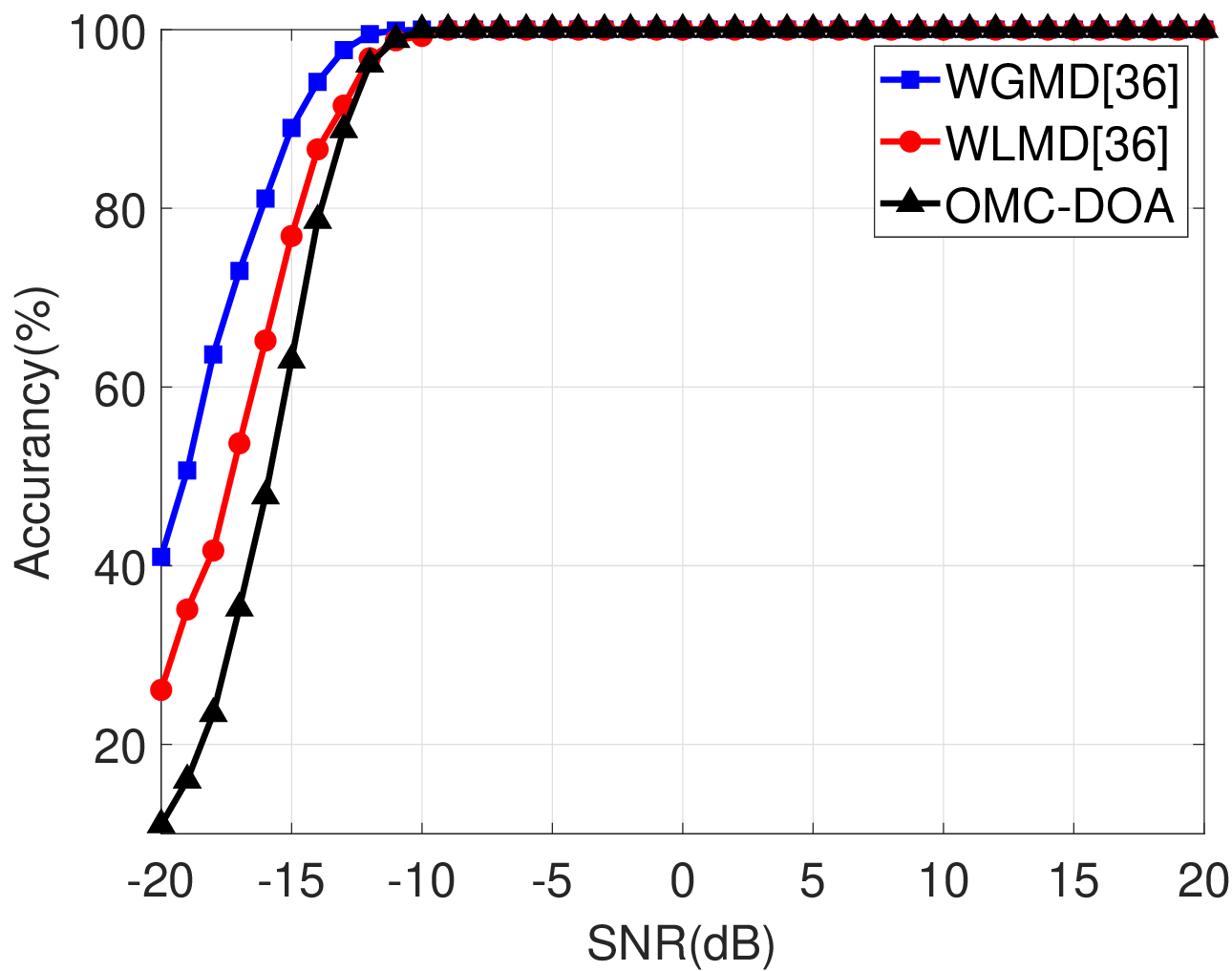}
        \caption{$\theta=11^{\circ}$}
        \label{Fig_accurancy1}
    \end{subfigure}
    \vspace{0.5em}  
    \begin{subfigure}{0.48\textwidth}
        \centering
        \includegraphics[width=\textwidth]{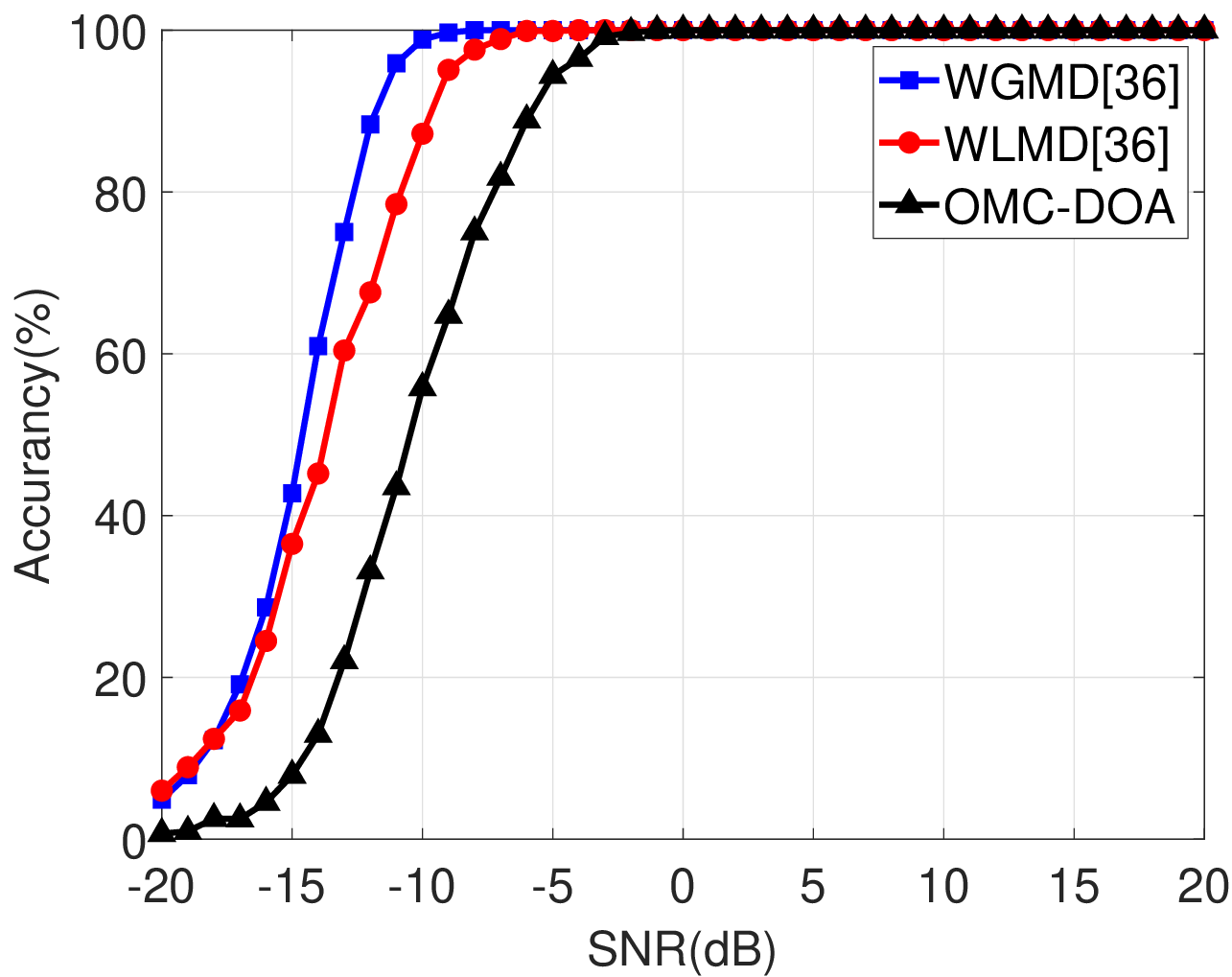}
        \caption{$\theta=23^{\circ}$}
        \label{Fig_accurancy2}
    \end{subfigure} 
    \caption{Accuracy versus SNR for DOA estimation.}
    \label{Fig_accurancy}
\end{figure}

Based on the RMSE results shown in Fig.~\ref{Fig_rmse_all}, Fig.~\ref{Fig_accurancy} presents the relationship between SNR and multi-source DOA estimation accuracy under the same simulation settings. 
Overall, the accuracy trends of the three methods are generally consistent with their RMSE performance. WGMD achieves 100\% accuracy at approximately -12 dB for $\theta_1 = 11^\circ$, while WLMD reaches this level slightly later. The OMC-DOA method converges more slowly in both directions, particularly for $\theta_2 = 23^\circ$, where a higher SNR is required for full identification. This phenomenon is related to the resolution differences inherent in the ESPRIT-generated candidate angles and also reflects the sensitivity of the OMC-DOA method to uneven candidate distributions. Nevertheless, all three methods achieve stable identification at medium-to-high SNRs, reaching 100\% accuracy and effectively accomplishing the multi-source DOA estimation task.
Also, the accuracy curves reach 100\% at a lower SNR than the point where the RMSE curves begin to level off in Fig.~\ref{Fig_rmse_all}. This is because the accuracy metric only reflects whether the true direction is correctly identified, while the RMSE reflects the magnitude of estimation error. Therefore, only when the estimation error further converges will the RMSE trend stabilize.

\begin{figure}[htp]
	\centering
    \includegraphics[width=0.48\textwidth]{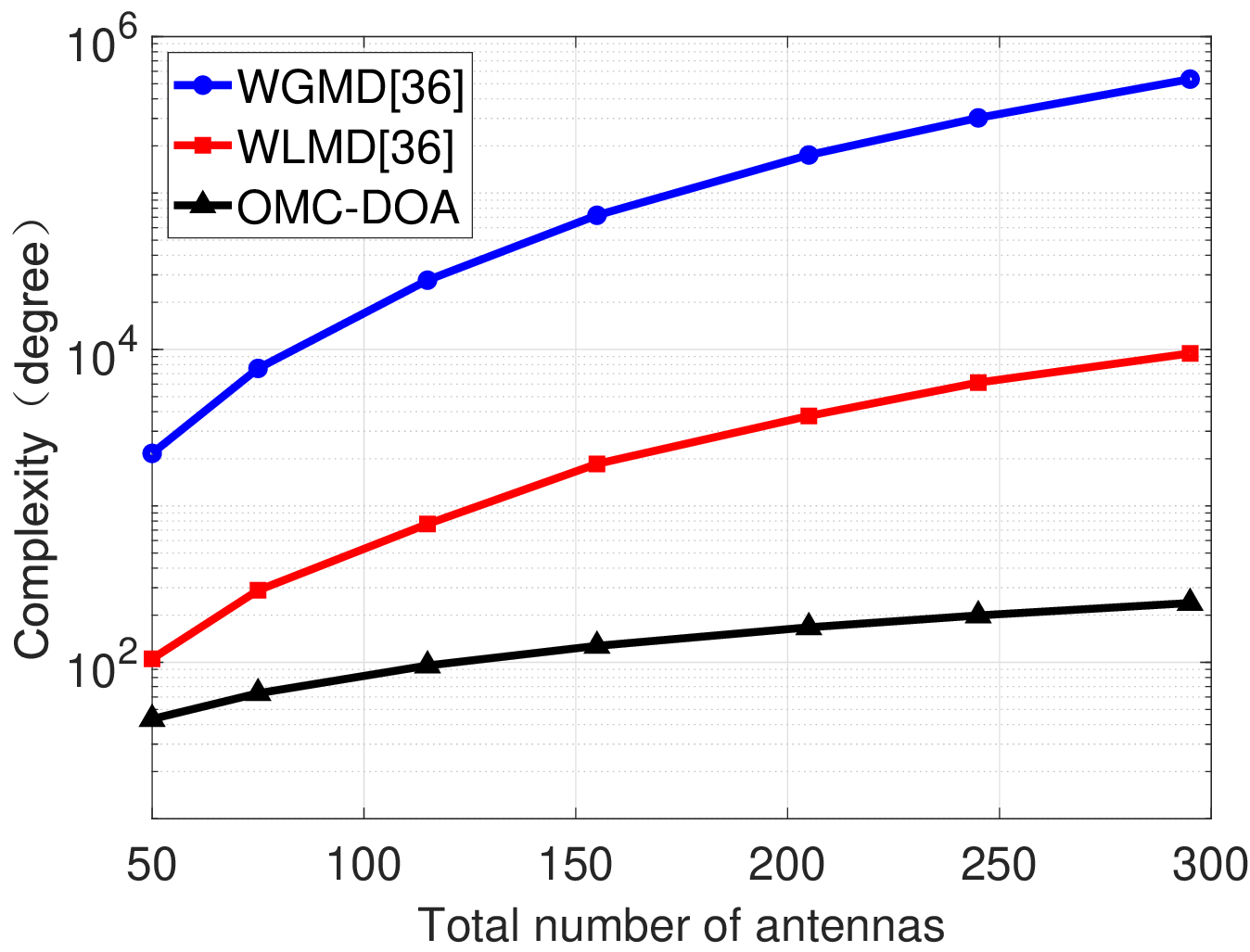}\\
	\caption{Complexity versus the number of antennas for DOA estimation.}\label{Fig_conplexity}
\end{figure}


Fig.~\ref{Fig_conplexity} illustrates the computational complexity under different numbers of antennas. Among them, WGMD consistently shows the highest complexity, which grows rapidly with the number of antennas due to its exhaustive search over all multi-subarray combinations and global optimization. WLMD exhibits moderate complexity, benefiting from its localized selection mechanism that reduces the search space among subarray candidates. In contrast, the OMC-DOA incurs the lowest computational burden, thanks to its streaming updates and lightweight clustering, which avoid centralized processing of the full candidate set and enhance scalability and real-time performance.
This comparison further validates the trade-off among the three methods in terms of computational resource consumption versus estimation accuracy.
The WGMD is preferable in high-accuracy scenarios but incurs significantly higher complexity, whereas the OMC-DOA, while slightly inferior in accuracy, is more suitable for resource-constrained environments or systems with strict real-time processing requirements.

%


\section{Conclusion}\label{sec:6}
This paper addressed the joint sensing of the number and directions of targets under the $\text{H$\mathrm{^2}$AD}$ MIMO architecture and proposed a two-stage sensing framework. In multi-source DOA estimation, accurate estimation of the number of sources was a critical prerequisite, as it significantly affected the performance of subsequent direction estimation. To this end, three source number sensing methods were designed: improved EDC, enhanced DNN incorporating statistical features and spectral entropy, and improved 1D-CNN based on the full eigenvalue sequence. Among them, the enhanced DNN and improved 1D-CNN mitigated the performance degradation of improved EDC under extremely low SNR conditions, where signal and noise eigenvalues became less distinguishable and clustering boundaries became ambiguous, thus enhancing the reliability of source number estimation. Based on the estimated number of sources, a low-complexity and high-accuracy DOA estimation method, named OMC-DOA, was further proposed. This method effectively reduced computational complexity and improved DOA estimation accuracy. In addition, a closed-form CRLB expression for DOA estimation under multi-source $\text{H$\mathrm{^2}$AD}$ conditions was derived. It was also shown that the steering vectors corresponding to different directions became asymptotically orthogonal in ultra-massive antenna systems, thereby further improving angular resolution. Simulation results demonstrated that the improved EDC, enhanced DNN, and improved 1D-CNN methods all achieved 100\% target number sensing accuracy for SNR $\ge$ –14 dB, with improved 1D-CNN maintaining superior performance even under extremely low SNR conditions. Meanwhile, the OMC-DOA method exhibited outstanding DOA sensing performance in medium-to-high SNR regions. Based on these findings, the proposed framework and methods show strong potential for applications in integrated sensing and communications and ultra-massive MIMO systems envisioned for future 6G networks.

\begin{appendices}
	\section{The derivations of CRLB.}\label{appendices}
As shown in~\cite{2009Classical}, when the SNRs of the individual signals are given and equal, and the number of signals is $A$, the FIM is an $A \times A$ matrix, i.e.,
\begin{align}
\mathbf{FIM} = 
\begin{bmatrix}
\mathrm{FIM}_{1,1} & \mathrm{FIM}_{1,2} & \cdots & \mathrm{FIM}_{1,A} \\
\mathrm{FIM}_{2,1} & \mathrm{FIM}_{2,2} & \cdots & \mathrm{FIM}_{2,A} \\
\vdots    & \vdots    & \ddots & \vdots \\
\mathrm{FIM}_{A,1} & \mathrm{FIM}_{A,2} & \cdots & \mathrm{FIM}_{A,A}
\end{bmatrix}.
\end{align}
The element in the $p$-th row and $r$-th column of the FIM can be given by
\begin{equation}
\mathbf{FIM}_{p,r} = \mathbf{Tr} \left( \mathbf{R}_y^{-1} \frac{\partial \mathbf{R}_y}{\partial \theta_p} \, \mathbf{R}_y^{-1} \frac{\partial \mathbf{R}_y}{\partial \theta_r} \right).
\end{equation}
Then, the signal received can be denoted as
\begin{equation}
\mathbf{y} = \mathbf{B}_A^H \mathbf{A}_s \mathbf{S} + \mathbf{w},
\end{equation}
where the beamforming matrix $\mathbf{B}_A$ and the array manifold matrix $\mathbf{A}_s$ are given as
\begin{equation}
\mathbf{B}_A = 
\begin{bmatrix}
\mathbf{B}_{A,1} & 0               & \cdots & 0 \\
0               & \mathbf{B}_{A,2} & \cdots & 0 \\
\vdots          & \vdots          & \ddots & \vdots \\
0               & 0               & \cdots & \mathbf{B}_{A,Q}
\end{bmatrix},
\end{equation}

\begin{equation}
\mathbf{A}_s = 
\begin{bmatrix}
\mathbf{a}_{1,1} & \cdots & \mathbf{a}_{1,A} \\
\vdots & \ddots & \vdots \\
\mathbf{a}_{Q,1} & \cdots & \mathbf{a}_{Q,A}
\end{bmatrix},
\end{equation}
respectively. The element $a_{q,i}$ in $\textbf{A}_s$ can be given by
\begin{equation}
 \mathbf{a}_{q,i} = (e^{j \frac{2\pi}{\lambda} d \sin \theta_i \sum_{qj=0}^{P - 1} N_{qj}} ) *
\begin{bmatrix}
1,
\cdots,
e^{j \frac{2\pi}{\lambda} (N_q - 1)d \sin\theta_i}
\end{bmatrix}^T.
\end{equation}
The covariance matrix $ \mathbf{R}_y $ of the received signal can be represented as
\begin{align}\label{43}
\mathbf{R}_y &= \mathbb{E}[\mathbf{y} \mathbf{y}^H]\notag\\ &= \mathbf{B}_A^H \mathbf{A}_s \mathbf{\Sigma}_s \mathbf{A}_s^H \mathbf{B}_A + \mathbf{I},\\
& = \sigma^2 \sum_{i=1}^{A}  \mathbf{B}_A^H 
\begin{bmatrix}
	\mathbf{a}_{1,i} \\
	\vdots \\
	\mathbf{a}_{Q,i}
\end{bmatrix}
\begin{bmatrix}
	\mathbf{a}_{1,i}^H & \cdots & \mathbf{a}_{Q,i}^H
\end{bmatrix}
\mathbf{B}_A + \mathbf{I}.\notag
\end{align}
where $\mathbf{\Sigma}_s =  \sigma^2 \mathbf{I}_A$, indicating that all signals share equal power. 
Expanding \eqref{43} yields~\eqref{eq:Ry}, as illustrated at the top of this page. The element $\textbf{R}_{y,q}$ in~\eqref{eq:Ry} can be expressed as

\begin{figure*}[!t]
    \centering
    \begin{equation}
    \label{eq:Ry}
    \begin{aligned}
    \textbf{R}_y &= \sigma^2 \sum_{i=1}^{A} 
    \begin{bmatrix}
    \mathbf{B}_{A,1}^H & 0 & \cdots & 0 \\
    0 & \mathbf{B}_{A,2}^H & \cdots & 0 \\
    \vdots & \vdots & \ddots & \vdots \\
    0 & 0 & \cdots & \mathbf{B}_{A,Q}^H
    \end{bmatrix}
    \begin{bmatrix}
    \mathbf{a}_{1,i} \\
    \vdots \\
    \mathbf{a}_{Q,i}
    \end{bmatrix}
    \begin{bmatrix}
    \mathbf{a}_{1,i}^H & \cdots & \mathbf{a}_{Q,i}^H
    \end{bmatrix}
    \begin{bmatrix}
    \mathbf{B}_{A,1} & 0 & \cdots & 0 \\
    0 & \mathbf{B}_{A,2} & \cdots & 0 \\
    \vdots & \vdots & \ddots & \vdots \\
    0 & 0 & \cdots & \mathbf{B}_{A,Q}
    \end{bmatrix}
    + \mathbf{I}\\
    &= \sigma^2\sum_{i=1}^{A} 
    \begin{bmatrix}
    \mathbf{B}_{A,1}^H \mathbf{a}_{1,i} \mathbf{a}_{1,i}^H \mathbf{B}_{A,1} & 0 & \cdots & 0 \\
    0 & \mathbf{B}_{A,2}^H \mathbf{a}_{2,i} \mathbf{a}_{2,i}^H \mathbf{B}_{A,2} & \cdots & 0 \\
    \vdots & \vdots & \ddots & \vdots \\
    0 & 0 & \cdots & \mathbf{B}_{A,Q}^H \mathbf{a}_{Q,i} \mathbf{a}_{Q,i}^H \mathbf{B}_{A,Q}
    \end{bmatrix} + \mathbf{I}
    =
    \begin{bmatrix}
    \mathbf{R}_{y,1} & 0 & \cdots & 0 \\
    0 & \mathbf{R}_{y,2} & \cdots & 0 \\
    \vdots & \vdots & \ddots & \vdots \\
    0 & 0 & \cdots & \mathbf{R}_{y,Q}
    \end{bmatrix}
    \end{aligned}
    \end{equation}\hrulefill
\end{figure*}
\begin{equation}
\begin{aligned}
\mathbf{R}_{y,q} &= \sigma^2\sum_{i=1}^{A}  \, \mathbf{B}_{A,q}^H \, \mathbf{a}_{q,i} \, \mathbf{a}_{q,i}^H \, \mathbf{B}_{A,q} +  \mathbf{I}_{M_q} .
\end{aligned}
\end{equation}

Thus, the $\text{FIM}_{p,r}$ under the $\text{H$\mathrm{^2}$AD}$ architecture can be given by
\begin{equation}\label{46}
\mathbf{FIM}_{p,r} = \sum_{q=1}^{Q} \mathbf{Tr} \left( \mathbf{R}_{y,q}^{-1} \frac{\partial \mathbf{R}_{y,q}}{\partial \theta_p} \, \mathbf{R}_{y,q}^{-1} \frac{\partial \mathbf{R}_{y,q}}{\partial \theta_r} \right),
\end{equation}
where
\begin{equation}\label{47}
\frac{\partial \mathbf{R}_{y,q}}{\partial \theta_i} 
= \sigma^2 \mathbf{B}_{A,q}^H \left( \dot{\mathbf{a}}_{q,i} \mathbf{a}_{q,i}^H + \mathbf{a}_{q,i} \dot{\mathbf{a}}_{q,i}^H \right) \mathbf{B}_{A,q}.
\end{equation}
Substituting \eqref{47} into \eqref{46} yields

\begin{equation}
\begin{aligned}
\MoveEqLeft[0] \hspace{2em}
\mathbf{Tr} \left( \mathbf{R}_{y,q}^{-1} \frac{\partial \mathbf{R}_{y,q}}{\partial \theta_p} \, \mathbf{R}_{y,q}^{-1} \frac{\partial \mathbf{R}_{y,q}}{\partial \theta_r} \right) \\
&= \sigma^4 \left(
\underbrace{
\mathbf{Tr}\left( \mathbf{R}_{y,q}^{-1} \mathbf{B}_{A,q}^H \dot{\mathbf{a}}_{q,p} \mathbf{a}_{q,p}^H 
\mathbf{B}_{A,q} \mathbf{R}_{y,q}^{-1} \mathbf{B}_{A,q}^H 
\dot{\mathbf{a}}_{q,r} \mathbf{a}_{q,r}^H \mathbf{B}_{A,q} \right)}_{T_1} \right. \\
&\quad + 
\underbrace{
\mathbf{Tr}\left( \mathbf{R}_{y,q}^{-1} \mathbf{B}_{A,q}^H \dot{\mathbf{a}}_{q,p} \mathbf{a}_{q,p}^H 
\mathbf{B}_{A,q} \mathbf{R}_{y,q}^{-1} \mathbf{B}_{A,q}^H 
\mathbf{a}_{q,r} \dot{\mathbf{a}}_{q,r}^H \mathbf{B}_{A,q} \right)}_{T_2} \\
&\quad + 
\underbrace{
\mathbf{Tr}\left( \mathbf{R}_{y,q}^{-1} \mathbf{B}_{A,q}^H \mathbf{a}_{q,p} \dot{\mathbf{a}}_{q,p}^H 
\mathbf{B}_{A,q} \mathbf{R}_{y,q}^{-1} \mathbf{B}_{A,q}^H 
\dot{\mathbf{a}}_{q,r} \mathbf{a}_{q,r}^H \mathbf{B}_{A,q} \right)}_{T_3} \\
&\quad + 
\left. \underbrace{
\mathbf{Tr}\left( \mathbf{R}_{y,q}^{-1} \mathbf{B}_{A,q}^H \mathbf{a}_{q,p} \dot{\mathbf{a}}_{q,p}^H 
\mathbf{B}_{A,q} \mathbf{R}_{y,q}^{-1} \mathbf{B}_{A,q}^H 
\mathbf{a}_{q,r} \dot{\mathbf{a}}_{q,r}^H \mathbf{B}_{A,q} \right)}_{T_4} \right),
\end{aligned}
\end{equation}
where
\begin{equation}
\dot{\mathbf{a}}_{q,i} = \frac{\partial \mathbf{a}_{q,i}}{\partial \theta_i} = \mathrm{j} \frac{\lambda}{2\pi} \cos\theta_i  \mathbf{D}_q \mathbf{a}_{q,i},
\end{equation}
where $\textbf{D}_q$ denotes a diagonal matrix, which can be defined as
\begin{equation}
\mathbf{D}_q = \mathbf{I}_{K_q} \otimes \mathbf{D}_{A_q} + \mathbf{D}_{D_q} \otimes \mathbf{I}_{M_q},
\end{equation}
where
\begin{equation}
 \mathbf{D}_{A_q} =\left[
\begin{array}{cccc}
d_1 & 0 & \cdots & 0 \\
0 & d_2 & \cdots & 0 \\
\vdots & \vdots & \ddots & \vdots \\
0 & 0 & \cdots & d_{M_q}
\end{array}
\right],
\end{equation}
\begin{equation}
\mathbf{D}_{D_q}=\left[
\begin{array}{cccc}
0 & 0 & \cdots & 0 \\
0 & d_{M_q+1} & \cdots & 0 \\
\vdots & \vdots & \ddots & \vdots \\
0 & 0 & \cdots & d_{(K_q-1)M_q+1}
\end{array}
\right].
\end{equation}

For clarity, we introduce an orthogonality assumption for the direction manifold vectors of different signals within the same subarray: for any subarray $q$, when $p \neq r$, the array manifolds are assumed to be orthogonal, as given by
\begin{equation}
\label{eq:0}
\mathbf{a}_{q,p} \mathbf{a}_{q,r}^H  = 0 .
\end{equation}
\eqref{eq:0} holds approximately under the condition that the number of antennas tends to infinity. The proofs are given below. \eqref{eq:0} can be rewritten as
\begin{equation}
\frac{1}{N_q}\,a_{q,p}^H\,a_{q,r}
=\frac{1}{N_q}\sum_{m=0}^{N_q-1}
e^{j\frac{2\pi d}{\lambda}m(\sin\theta_r-\sin\theta_p)}.
\end{equation}
Here, we determine orthogonality by computing the normalized inner product. Define the phase difference caused by the angle difference as $\Delta = \tfrac{2\pi d}{\lambda}(\sin\theta_r - \sin\theta_p)$. Then, the inner product can be expressed as
\begin{equation}
\frac{1}{N_q}\sum_{m=0}^{N_q-1} e^{j m \Delta}.
\end{equation}
When $\theta_p = \theta_r$, $\Delta = 0$. We have
\begin{equation}\label{56}
\frac{1}{N_q}\sum_{m=0}^{N_q-1} e^{j m \cdot0} = \frac{1}{N_q}\sum_{m=0}^{N_q-1} 1 = 1.
\end{equation}
When $\theta_p \neq \theta_r$, $\Delta\neq 0 $. We have
\begin{equation}\label{57}
\frac{1}{N_q}\sum_{m=0}^{N_q-1} e^{j m \Delta}
=\frac{1}{N_q}\cdot\frac{1 - e^{j N_q \Delta}}{1 - e^{j\Delta}},
\end{equation}
As the number of antennas approaches infinity, i.e., $N_q \to \infty$, we can obtain
\begin{equation}\label{58}
\lim_{N_q\to\infty}\frac{1}{N_q}\cdot\frac{1 - e^{j N_q \Delta}}{1 - e^{j\Delta}}=0.
\end{equation}
According to \eqref{56}, \eqref{57}, and \eqref{58}, we have
\begin{equation}
\lim_{N_q\to\infty}\frac{1}{N_q}a_{q,p}^H a_{q,r}
=
\begin{cases}
1, & \theta_p=\theta_r,\\
0, & \theta_p\neq\theta_r.
\end{cases}.
\end{equation}

Base on~\eqref{eq:0}, when $p \neq r$, we can obtain
\begin{equation}
\mathbf{a}_{q,p} \mathbf{a}_{q,r}^H = \dot{\mathbf{a}}_{q,p} \mathbf{a}_{q,r}^H = \mathbf{a}_{q,p} \dot{\mathbf{a}}_{q,r}^H = 0 .
\end{equation}
Thus, all the cross terms vanish, resulting in $T_1 = T_2 = T_3 = T_4 =0$. The FIM becomes diagonal, and can be expressed as

\begin{equation}
\mathbf{FIM} = \mathrm{diag} ( \mathrm{FIM}_1, \mathrm{FIM}_2, \dots, \mathrm{FIM}_A ).
\end{equation}
The FIM $\mathrm{FIM}_i$ corresponding to the $i$-th signal source can be given by
\begin{equation}
\mathrm{FIM}_i = \sum_{q=1}^{Q} \mathbf{Tr} \left( \mathbf{R}_{y,q}^{-1} \frac{\partial \mathbf{R}_{y,q}}{\partial \theta_i} \, \mathbf{R}_{y,q}^{-1} \frac{\partial \mathbf{R}_{y,q}}{\partial \theta_i} \right).
\end{equation}
On the basis of the above analysis, the CRLB for multi-souce DOA sensing is repressed as 
\begin{equation}
\mathrm{CRLB}(\theta_i) = \frac{1}{L}[\mathrm{FIM}^{-1}]_{i,i} = \frac{1}{L \cdot \mathrm{FIM}_i} ,
\end{equation}
where $L$ denotes the number of independent snapshots. Therefore, the CRLB can be deduced as
\begin{equation}
\mathbf{CRLB} = \mathrm{diag} \left\{ 
\frac{1}{L \cdot \mathrm{FIM}_1},\ 
\frac{1}{L \cdot \mathrm{FIM}_2},\ 
\dots,\ 
\frac{1}{L \cdot \mathrm{FIM}_A} 
\right\}.
\end{equation}
\end{appendices}










\renewcommand\refname{References}
\bibliographystyle{IEEEtran}
\bibliography{cite}

\end{document}